\documentclass{article}
\usepackage{arxiv}
\usepackage[utf8]{inputenc} % allow utf-8 input
\usepackage[T1]{fontenc}    % use 8-bit T1 fonts
\usepackage{hyperref}       % hyperlinks
\usepackage{url}            % simple URL typesetting
\usepackage{booktabs}       % professional-quality tables
\usepackage{amsfonts}       % blackboard math symbols
\usepackage{amsmath}
\usepackage{nicefrac}       % compact symbols for 1/2, etc.
\usepackage{microtype}      % microtypography
\usepackage{cleveref}       % smart cross-referencing
\usepackage{lipsum}         % Can be removed after putting your text content
\usepackage{graphicx}
\usepackage{natbib}
\usepackage{doi}
\usepackage[dvipsnames]{xcolor}
\usepackage{graphicx}
\usepackage{graphbox}
\usepackage{setspace}
\usepackage{lscape}
\usepackage{rotating}
\usepackage{adjustbox}
\usepackage{makecell}
\usepackage{threeparttable}
\usepackage{caption}
\usepackage{csquotes}

\newcommand{\blandscape}{\begin{landscape}}
\newcommand{\elandscape}{\end{landscape}}

\title{Credible causal inference beyond toy models}
\date{February 18, 2024}

\newif\ifuniqueAffiliation
\uniqueAffiliationtrue

\ifuniqueAffiliation % Standard variant of author block
\author{{Pablo Geraldo Bastías} \\
	Nuffield College\\
	University of Oxford\\
	\texttt{pablo.geraldo@nuffield.ox.ac.uk}}

\renewcommand{\headeright}{}
\renewcommand{\undertitle}{An invitation to graphical models for the skeptic}
\renewcommand{\shorttitle}{Credible causal inference beyond toy models}

\hypersetup{
pdftitle={Credible causal inference beyond toy models},
pdfsubject={},
pdfauthor={Pablo Geraldo Bastías},
pdfkeywords={causal inference, DAG, quasi-experiments},
}

\begin{document}
\maketitle

\begin{abstract}
Causal inference with observational data critically relies on untestable
and extra-statistical assumptions that have (sometimes) testable
implications. Well-known sets of assumptions that are sufficient to
justify the causal interpretation of certain estimators are called
\emph{identification strategies}. These templates for causal analysis,
however, do not perfectly map into empirical research practice.
Researchers are often left in the disjunctive of either abstracting away
from their particular setting to fit in the templates, risking erroneous
inferences, or avoiding situations in which the templates cannot be
applied, missing valuable opportunities for conducting empirical
analysis. In this article, I show how directed acyclic graphs (DAGs) can
help researchers to conduct empirical research and assess the quality of
evidence without excessively relying on research templates. First, I
offer a concise introduction to causal inference frameworks. Then I
survey the arguments in the methodological literature in favor of using
research templates, while either avoiding or limiting the use of causal
graphical models. Third, I discuss the problems with the template model,
arguing for a more flexible approach to DAGs that helps illuminating
common problems in empirical settings and improving the credibility of
causal claims. I demonstrate this approach in a series of worked
examples, showing the gap between identification strategies as invoked
by researchers and their actual applications. Finally, I conclude
highlighting the benefits that routinely incorporating causal graphical
models in our scientific discussions would have in terms of
transparency, testability, and generativity.
\end{abstract}

% keywords can be removed

\section{Introduction}

It is not new that social scientists are interested in answering causal
questions. What is new is the way they deal with the challenges of doing
so using non-experimental data. Traditionally, concerns about biases
arising from ``confounding factors'' were addressed using all sorts of
regression models, with the inclusion of control variables motivated on
more or less informal grounds. While, perhaps, the majority of published
quantitative research still fits this description, the practice is
rapidly falling out of favor. In its place, researchers are moving
towards what has been called a design-based approach to causal
inference, characterized by an emphasis on defining the target quantity
outside of any statistical model \citep{LundbergWhat2021}, a focus on
the (usually non-parametric) assumptions needed to identify such
quantities with observational data, and a reliance on quasi-experimental
studies. This approach is dramatically transforming quantitative
research, a process that has been described as the rise of ``causal
empiricism'' \citep{SamiiCausal2016}, a ``credibility revolution''
\citep{AngristCredibility2010}, or simply a ``causal revolution''
\citep{PearlBook2018}.

It is widely accepted that to move from associations to causal relations
one needs to make assumptions about the data generating process that
cannot be verified in the joint probability distribution of the
observable variables of interest. In other words, causal inference
requires \emph{untestable} identification assumptions involving \emph{by
definition} unobservable quantities \citep{BareinboimPearl2020} or, as
concisely stated by Nancy Cartwright, ``no causes in, no causes out''
\citep{CartwrightNo2020, CartwrightPrecis1995}. However, one thing is
being able to make the assumptions explicit; judging if they are
plausible \emph{in a given application} is an entirely different
challenge, and a major source of disagreement across research
communities. In fact, it is a real risk to simply show a laundry list of
identification assumptions without careful examination of their
plausibility, or without conducting appropriate sensitivity analyses to
potential departures from perfect compliance of the identifying
assumptions\footnote{See, for example, the discussion about matching methods in \citet{SekhonOpiates2009}. The general problem is, of course, not limited to matching or selection on observables assumptions.}.
Paraphrasing \citet{PetersenCausal2014}, many times researchers rely on
convenience-based rather than knowledge-based assumptions.

Probably in part as a reaction to this, social scientists have turned to
so-called quasi-experiments, trying to find settings that can credibly
approximate an experimental ideal, in which identification assumptions
hold by design. Quasi-experiments have become the preferred type of
identification strategy, even the only one worth referring as ``research
design''. The logic of the argument is as follows: quasi-experiments are
more credible because, by exploiting some well-understood variation in
the exposure of interest, they bypass the need of understanding and
modeling the entire data generating process. Only a few templates meet
this condition. On the contrary, according to this view, using graphical
models assumes that researchers are capable of accurately modeling the
relevant processes causing both the treatment and the outcome of
interest, an impossibly high bar, at least for social scientists.
Therefore, it would be better to avoid using graphical models
altogether.

In this article, I argue that, contrary to common wisdom, the utility of
graphical models is not necessarily opposed to quasi-experimental or
design-based approaches. It is true that not many areas of social
research, if any, are simple enough for researchers to start writing a
model top-down or deductively, guided only by theory and available
evidence. It is also true that most aspects of social reality are
unknown or very partially understood, and this is why researchers
interested in causal inference have turned to more ``controlled''
environments to test their hypotheses. Yet the choice between
\emph{model-free} as opposed to \emph{model-based} causal inference is a
false dilemma, diverting us from addressing the mostly harmful issue of
\emph{model-blind} research. If social sciences are to advance our
understanding of the social world based on causally grounded evidence,
neither a general or unrestricted causal models has to be the starting
point, nor crumbs of disconnected evidence should be the final step.
Models, as encoded by graphical or other means, are not to be feared;
they are but a useful scaffolding in the dialogue between theory and
evidence.

In what follows, I make two simple claims. First, actual research do not
unambiguously map into well known templates or research designs
(\emph{identification strategies}); instead, it usually requires a fuzzy
mix of assumptions that go beyond the labels that authors use to
classify (and, many times, to signal) the credibility of their own work.
Second, graphical models can help us to conduct better research in messy
settings and to asses the strength of evidence \emph{as it is}, making
the assumptions required for a study to be causally interpreted more
transparent, without relying on excessively rigid templates. I aim to
show that doing so do not imply imposing stronger assumptions than those
social scientists are routinely willing to make, but only to express
them in a way that is in general more intelligible. In short, my goal is
to help researchers to make better judgments about the credibility of
causal claims with the aid of graphical models.

To develop this argument, I take the following steps. First, I offer a
concise introduction to the most widely used causal inference
frameworks, focusing on how they formulate and solve the identification
problem. Then I survey the arguments in the methodological literature in
favor of using research templates, while either limiting or avoiding the
use of causal graphical models. Third, I address the question of where
do DAGs come from, arguing that, in general, current practice falls
short in exploiting the full expressiveness of graphical models. I then
propose a minimalistic or agnostic approach to using graphical models in
causal inference, that I believe can help us to move the discussion
about credibility beyond research templates. Through a series of
examples using causal diagrams to encode the substantive assumptions
implicitly made by researchers in their empirical work, I demonstrate
the gap between applications and the identification strategies they
invoke. Finally, I provide recommendations for practitioners on how to
assess the credibility of causal arguments to be more sensible to a
study's particular setting, combining testable implications and
sensitivity analysis.

\section{Frameworks for causal inference: A side-by-side introduction}

Two conceptual frameworks dominate the causal inference literature:
potential outcomes (PO, a.k.a. Neyman-Rubin causal model) and the
structural causal model (SCM, a.k.a. Pearl's graphical approach). Mostly
for idiosyncratic and disciplinary reasons they have been independently
adopted in different research communities. In statistics and economics,
potential outcomes remain the uncontested hegemonic
language\footnote{In the case of economics, I am mostly referring to current practice in empirical microeconomics, favouring a mix of potential outcomes and quasi-experiments. There is, of course, a long tradition of structural estimation in economics, which is closely related to the development of modern, non-parametric structural causal models.},
while in computer science and epidemiology the SCM has become the
community's \emph{koine}.

Although not everyone agrees on the necessity of having a specific
language tailored to answer causal questions
\citep{MaclarenWhat2020,DawidCausal2000,DawidDecisiontheoretic2020},
proponents of PO and SCM have highlighted the specificity of causal
inference in relation to traditional statistics and, more recently,
machine learning (see Table \ref{tab:causality_ladder} for reference).
Both statistics and machine learning excel at describing and making
inferences (or predictions) about a data generating process that is, in
general, being passively observed. The usual theoretical quantities a
statistical method is trying to approximate (their \emph{estimand}) are
functions of, in principle, observable joint probability distributions.
Maybe the most prominent example for social scientists would be
estimating the expected value of an outcome \(Y\) given the value of a
predictor variable \(X\): \(E(Y|X=x)\), as it is done in traditional
regression models. The type of questions one is able to address in these
terms are those related to the association between variables, telling us
how to update our beliefs when new information becomes available.

\begin{table}[h]
\begin{threeparttable}
\caption{The Ladder of Causality (adapted from \cite{PearlBook2018})}
\label{tab:causality_ladder}
\begin{tabular}{l | l l l}
\toprule
& Association & Intervention & Counterfactuals\\
\midrule
Estimand & $\mathbf{P(Y \vert X)}$ & $\mathbf{P(Y \vert do(x))}$ & $\mathbf{P(Y_x \vert x',y')}$\\
\\
Activity & Seeing, Observing & Doing, Intervening & Imagining, Retrospecting \\
\\
Fields & \makecell[cl]{Statistics\\Machine Learning} & \makecell[cl]{Experiments\\Policy evaluation} & \makecell[cl]{Structural models}\\
\\
Questions & \makecell[cl]{What would I believe \\about Y if I see X?} & \makecell[cl]{What would happen \\with Y if I do X?} & \makecell[cl]{What would have happened \\with Y have I done X \\instead of X'? Why?}\\
\\
Example & \makecell[cl]{What is the expected income \\of a typical college graduate?} & \makecell[cl]{What would be the income \\of a college graduate, \\if everyone attend college?} & \makecell[cl]{What would be my income\\ have I graduated from college \\given that I didn't attend?}\\
\toprule
\end{tabular}
\end{threeparttable}
\end{table}

In contrast, causal inference focus on making predictions under
(potentially hypothetical) interventions on the data generating process,
not merely a passive observation of it. Conceptually, the appropriate
estimand for this scenario involves potential outcomes, the what-if
outcomes an individual (or unit) would experience under an intervention
that \emph{assigns} a given value to an exposure variable (instead of
just recording its naturally occurring value), while leaving everything
else (not affected by the exposure) unmodified. Note that, in general,
those potential outcomes would \emph{not} be equal to the conditional
expectation of \(Y\) given \(X\), a problem that researchers have long
recognized and cataloged under the umbrella terms of self-selection,
confounding and omitted variable bias.

Contemporary causal inference unify these issues under a general
framework, making explicit that the identification of causal effects
from observational data requires \emph{causal assumptions}, i.e.,
assumptions about the data generating process that are
extra-statistical, in the sense that they cannot be strictly verified or
tested looking at the observed distributions, assumptions that would be
required even if we have an infinite number of observations. In this
context, \emph{causal identification} refers to the task of
re-expressing an interventional distribution in terms of observables
alone, using \emph{causal assumptions}.

In what follows, I briefly introduce the aforementioned causal inference
frameworks, emphasizing their particularities in how the identification
task is formulated, and their logical equivalence, including how to move
from one formalization to the other. More detailed introductions to
potential outcomes can be found in
\citet{ImbensCausal2015, RubinCausal2005, Rubinobjective2008}, and to
the structural causal model in
\citet{PearlCausal2016, PearlCausality2009, PearlFoundations2010}.
Readers familiar with any of these frameworks can skip the corresponding
section, although they might still benefit from the particular focus of
this concise introduction.

\subsection{Potential outcomes (PO)}

\subsubsection{Defining potential outcomes}

In the potential outcomes framework, we usually start with a unit of
interest, let's say, a particular individual \(i\). As a running
example, let's suppose that we want to quantify the returns of
education. More precisely, we want to know if attending college
\emph{causes} the income of a person to be higher than it would have
been has this person not attended college. By assumption, and before
seeing any data, we suppose that this person have some \emph{potential
outcomes}, i.e., hypothetical income levels indexed by their educational
level. Denoting income as \(Y\) and education as \(D\), with
\(D \in \{0,1\}\), where \(D=0\) indicates no college education and
\(D=1\) indicates college education, we can write individual \(i\)'s
income have they attended college as \(Y_{d=1, i} = Y_{1i}\), and their
income have they not attended college as \(Y_{d=0, i} = Y_{0i}\).

We say that \(D\) has a causal effect on \(Y\) if
\(Y_{1i} \neq Y_{0i}\). In general, a given \emph{causal estimand} or
causal effect would be defined as a contrast of potential outcomes. For
example, we can define \(\tau_i = Y_{1i} - Y_{0i}\) as the
\emph{individual treatment effect} (ITE), but other contrasts (such as
ratios) are certainly possible. Keep in mind that, in the potential
outcomes framework, this individual effect will be our building block
for other \emph{average} effects, taking the expectation of \(\tau_i\)
over a given group or population (see \cite{LundbergWhat2021} for a more
detailed discussion).

Notice that, just by formalizing the potential outcomes in this way, we
already introduced an assumption of \emph{consistency} that is baked in
the notation itself. This assumption, often referred as the Stable Unit
Treatment Value Assumption or SUTVA, indicates that individual \(i\)'s
potential outcomes respond solely to its own exposure level. This
assumption would not hold if, for example, the potential outcomes of one
unit depend on the exposure of other units. Going back to the question
of returns of education, this would be the case if your income as a
college graduate depends on your family's and friends' educational
attainment (network effects), or depends on the proportion of college
graduates in the population (general equilibrium effects). By writing
only \(Y_{di}\), we discard such
effects\footnote{A different issue that would also imply a threat to the validity of the consistency assumption is when treatments or exposures are not \textit{well-defined}, in the sense that $D$ could mean different things. In that case, $Y_d$ would be an unspecific and ill-defined quantity. See \citet{HernanCausal2023} for more details.}.

The next step is to link potential to observed outcomes, and we do this
by assumption too. Following our example, let's denote the individual's
\emph{observed} outcome, i.e., their actual income, as \(Y_i\), and
their \emph{observed} exposure, i.e., their attained education level, as
\(D_i\). By \emph{consistency} (again), we assume that the observed
outcome corresponds to the potential outcome under the treatment level
actually received
(\(D_i=d \rightarrow Y_i = Y_{di}\))\footnote{In the binary case, this can be written as $Y_i = Y_{1i}(D_i)+Y_{0i}(1-D_i)$, sometimes referred as the``switching equation'' for potential outcomes.}.
Given that only one treatment, and therefore only one potential outcome
can be realized for each \(i\), we can never directly calculate
\(\tau_i\). We get to observe at most one potential outcome
(corresponding to \(Y_i\)), while at least one potential outcome remains
\emph{counterfactual}\footnote{In the PO framework, it is common to refer to both potential outcomes as \textit{counterfactuals}. However, this can be misleading both because as it was just mentioned only some potential outcomes remain counterfactual, and also because the concept has the stronger meaning of the outcome under an exposure \textit{contrary} to the one that actually happened, a situation that arises for example in causal attribution.}.
The impossibility of observing both potential outcomes simultaneously
was described as the \emph{fundamental problem of causal inference} by
\citet{HollandStatistics1986}.

\subsubsection{Estimand, Bias, and Identification using PO}

Given the impossibility of identifying individual treatment effects
without heroic assumptions, researchers usually resort to average
effects, taking the expectation of \(\tau_i\) over a group of units
under
study\footnote{For the sake of simplicity, here I am making abstraction of the difference between the sample and population. For a discussion on that important topic, see \citet{ImaiMisunderstandings2008}}.
For example, we can start by making our quantity of interest the Average
Treatment Effect (ATE), that we can define as:
\(E(\tau_i) = E(Y_{1i} - Y_{0i}) = E(Y_{1i}) - E(Y_{0i})\)

Where the last equality holds by linearity in expectations. In our
example, the ATE correspond to the difference in average income in the
population, have all the units attained college education, versus none
of them having college education. Other common \emph{causal estimands}
are the Average Treatment Effect on the Treated
(\(ATT = E[Y_{1i} - Y_{0i} | D_i = 1]\)), the Average Treatment Effect
on the Untreated or Controls (\(ATC = E[Y_{1i} - Y_{0i} | D_i = 0]\)),
and the Conditional Average Treatment Effect
(\(CATE = E[Y_{1i} - Y_{0i} | X_i = x]\)), where \(X_i\) denotes
observed covariate(s), such as age, race, gender,
etcetera\footnote{A conditional effect is any effect estimated on a subpopulation, therefore both ATT and ATC are conditional effects too. The Local Average Treatment Effect (LATE) estimated in the instrumental variable framework under monotonicity is also a conditional effect. Following the convention, I reserve CATE to denote conditioning on observed covariates.}.

Notice that these average effects are still theoretical estimands, and
they cannot be computed from data without further assumptions. So the
next step would be to recognize which \emph{identification assumptions}
allow us to rewrite our desired estimands in terms of purely
observational quantities. These conditions can take the form of
(conditional) independences between potential outcomes and observed
variables, and/or parametric assumptions about the functional form of
the potential outcomes, including linearity, continuity, monotonicity,
etcetera.

In practice, a common way to reveal which assumptions one would need in
order to identify a causal effect with observational data is to start
with some function of the observed data (an \emph{estimator}) that
researches might consider to use as an approximation to their quantity
of interest (an \emph{estimand}), trying to rewrite its expectation in
terms of potential outcomes. By doing so, some undesirable terms will
appear that make the expectation of our estimator different from our
quantity of interest (these are sources of \emph{bias}). We call
\emph{identification assumptions} to the conditions needed to get rid of
those additional terms and, therefore, to establish the equality between
the expectation of our estimator and our estimand. Let's see, for
example, what happen when we want to estimate the ATE using a
difference-in-means estimator, a natural choice for a two-groups
contrast. After some algebra, we will arrive to some well-known
decompositions such
as\footnote{See \citet{MorganCounterfactuals2014} for full derivations.}:

\begin{equation*}
\begin{aligned}
\mathbb{E}[\text{diff-in-means}] = \mathbb{E}[Y_1|D_i=1] - \mathbb{E}[Y_0|D_i=0] \\
= \underbrace{\mathbb{E}[Y_{1i} - Y_{0i}]}_{\text{ATE}}  
+ \underbrace{(1 - P[D_i])}_{\text{Control weight}} ~ \underbrace{(\mathbb{E}[Y_{1i}|D_i=1] - \color{red}{\mathbb{E}[Y_{1i}|D_i=0]}\color{black}{)}}_{\text{Bias from control units' unobserved POs}}
+ \underbrace{(P[D_i])}_{\text{Treated weight}} ~ \underbrace{(\color{red}{\mathbb{E}[Y_{0i}|D_i=1]}\color{black}{-\mathbb{E}[Y_{0i}|D_i=0])}}_{\text{Bias from treated units' unbserved POs}}\\
\end{aligned}
\end{equation*}

Where the quantity in the first line is the expectation of the
difference-in-means estimator written as a function of purely observed
outcomes, and the following line show that this is equivalent to the
true ATE (\(\mathbb{E}[Y_{1i} - Y_{0i}]\)) plus some bias terms: the
expected difference in the treated \emph{potential} outcomes (\(Y_1\))
between \emph{actually} treated and untreated units (weighted by the
probability of being untreated), plus the expected difference in
untreated \emph{potential} outcomes between units \emph{actually}
receiving treatment and control (weighted by the probability of being
treated). The terms in red correspond to those that are not observable,
and therefore the focus of our
assumptions\footnote{We can further rearrange the terms to obtain another useful decomposition: $$\mathbb{E}[\text{diff-in-means}] = \mathbb{E}[Y_1|D_i=1] - \mathbb{E}[Y_0|D_i=0] \\ = \mathbb{E}[Y_{1i} - Y_{0i}] + (\mathbb{E}[Y_0|D_i=1] - \mathbb{E}[Y_0|D_i=0]) + (1-P[D])(ATT-ATC)$$ In words, this imply that the difference-in-means estimator is equal to the true ATE, plus a baseline bias term (the difference in control potential outcomes between treated and controls units), plus the differential response to the treatment between treated and control units (the ATT minus the ATC, weighted by the probability of being control).}.

With this decomposition in hand, we can then ask: what assumption would
be necessary to get rid of the bias terms? Clearly, one admissible
answer is that there is no difference in the \emph{potential} outcomes
between \emph{actually} treated and untreated units. In others words, we
need to assume that the expectation of \(Y_d\) is independent of the
treatment status, formally denoted as
\(\{Y_{1i}, Y_{0i}\}\perp\!\!\!\perp D_i\). This assumption is known as
\emph{ignorability} or \emph{exchangeability}, because it implies that
we can ignore the conditioning on \(D_i\), and exchange the \(Y_d\)
between the treated and the untreated. In our example, this would imply
that the potential income of college graduates have they not attended
college is the same as the expected (observed) income of those who did
not attend college, and
viceversa\footnote{As it might be already evident, if instead of the ATE, we want to estimate the ATT or ATC, we need a weaker condition, known as \textit{partial exchangeability}, where only one of the potential outcomes is required to be independent of the treatment.}.

It is important to note that, in experimental settings, this conditions
holds \emph{by design}: being \emph{randomized}, the treatment
assignment vector \(D\) does not contain information whatsoever about
the potential outcomes. In observational settings, on the other hand, it
is difficult to immediately believe in such a convenient assumption. One
might want to argue, however, that it is likely more plausible that
those potential outcomes are exchangeable only between units that are
similar in terms of their observed covariates \(X_i\). The updated
assumption then becomes \emph{conditional} ignorability or
exchangeability. If this is the case, instead of using the simple
difference-in-means discussed above, we might want to use a conditional
estimator to obtain strata-specific or conditional expectations, and
then recover the unconditional difference averaging over
subgroups\footnote{We can additionally estimate these effects for the entire population or for a different target (sub)population, with appropriate weights. For example, if we want the ATE, we would use the distribution of $X$ for the entire population; however, if we want the ATT (or ATC), we should use as weights the distribution of $X$ for the treated (untreated) subpopulation.}.

But now it seems that we have solved one problem by creating two new
issues. First, if we are going to exchange potential outcome between
treated and untreated units conditional on some values of \(X\), we need
to have both treated and untreated units with the same (or similar
enough) values of \(X\), a condition known as \emph{positivity},
\emph{overlap} or \emph{common support}. Second, if we are assuming that
ignorability holds conditional on the values of certain \emph{observed
covariates}, we are left wondering: which covariates should be
conditioned on? Even more, do we observe all the relevant variables, or
are we missing some covariate(s) that, if not included, would make the
ignorability assumption false?

\subsubsection{Heuristics: Assignment Mechanism}

Up to this point, we have seen how useful potential outcomes are in
order to define what do we want to know (\emph{causal estimand}), to
clarify the sources of discrepancy between what we observe and what we
want (\emph{sources of bias}), and to formalize what needs to be true
for our estimand to be identified with a given estimator and
observational data (\emph{identification assumptions}). The next step
would be to assess the plausibility of the invoked assumptions. The
difficulty is that this requires assessing (in)dependence relationships
between counterfactual variables. Therefore, we need an indirect way of
evaluating the assumptions.

The heuristic approach developed, in the PO tradition, to deal with the
problem of what to condition for to render our setting experiment-like
is to understand and model the \emph{assignment mechanism}: the factors
that explain why some units end up exposed and some do not. Accounting
for these factors, all remaining variation in treatment status is
assumed to be not associated with the potential outcomes. Most
importantly, one should account for variables considered to be
\emph{confounders}: those jointly affecting the treatment and the
outcome. In our example of returns of education, they could be factors
like parental socioeconomic status, students academic ability, among
others. But also, following the very idea of the \emph{assignment
mechanism}, one might want to include variables that are known (or
believed) to affect the treatment status even if their relationship to
the outcome is
unknown\footnote{This is generally not a good strategy, since it can hurt efficiency, and if there is remaining confounding these variables can act as ``bias amplifiers'' \cite{CinelliCrash2022}.}.

The analogy with randomization can be clarifying. If the treatment is
assigned at random (i.e., with independence of the potential outcomes
and covariate values), we can directly estimate causal quantities from
the observed data, because unconditional ignorability
(\(Y_d \perp\!\!\!\perp D\)) holds \emph{by design}. Even if the
probability of receiving the treatment varies by some characteristics of
the units (as in the case of block experiments), knowing these
probabilities, i.e., a unit's \emph{propensity score}
\citep{Rosenbaumcentral1983} we can still recover the effects of
interest using a conditional estimator, because conditional ignorability
(\(Y_d \perp\!\!\!\perp D | X\)) holds again \emph{by design}. And even
if we don't know such probabilities but we (assume to) know which
observed variables they depend upon, we can estimate these
\emph{propensity scores} and recover the true effect, because
conditional ignorability holds (this time, \emph{by assumption}). In
other words, identification in observational setting is approached by
analogy or \emph{reductio ad aleatorium}: we ask ourselves, how
experiment-like is our setting? Can we possibly assume that the exposure
of interest is as-if random (maybe after conditioning for some
covariates)?

For these reasons, the advice has been in general to use all ``true
covariates'', i.e., all variables occurring pre-treatment and none of
the post-treatment variables
\citep{ImbensCausal2015, Rubindesign2007, Rubinobjective2008}. According
to this view, it is a good idea to include all the information
available; at worst, one would include irrelevant variables (affecting
efficiency), but this is preferable to the risk of omitting something
important (inducing
bias)\footnote{For a more detailed and up to date discussion of what constitutes good or bad controls, see \citet{CinelliCrash2022}.}.
Furthermore, with recent developments and the adoption of machine
learning methods in causal inference settings, it has been argued that
adjusting for high dimensional covariates can make the conditional
ignorability assumption more plausible
\citep{AtheyRecursive2016, EcklesBias2021}. Even in this case, however,
when dealing with observational data we cannot be sure that all relevant
variables have been included, so it is generally advised to implement
formal sensitivity analyses to assess how much our conclusions would
change in the presence of hidden bias
\citep{RosenbaumDesign2010, CinelliMaking2020, VanderWeeleBias2011}.

On the other hand, with the rise of quasi-experimental designs, the
emphasis has shifted towards finding settings where the as-if random
assumption hold more naturally, \emph{almost} ``by design'' as in true
experiments. The role of the researcher, instead of motivating a
(however informal) model of the joint relation between treatment and
outcome, would be to find situations where the assignment mechanism
(i.e., who receives the treatment and who does not), even if not under
the researchers' own control, obeys to simple rules (like an
administrative threshold) or responds to haphazard events (like a
natural disaster). By sticking to these simpler situations, the argument
goes, we can bypass the need to model a complex assignment mechanism,
increasing our confidence on not missing variables that, if omitted,
would bias our analyses. Generally, the price to pay for this solution
is a change in the estimand we are able to recover, which becomes more
``local'' in the sense of representing only a particular subpopulations
of units \citep{ImbensBetter2010a}.

In summary, due to its simplicity and experiment-like flavor, the
heuristic of the assignment mechanism results appealing for applied
researchers. At the same time, it seems excessively restrictive and
potentially misleading. It is restrictive because, as we will see,
having a treatment assigned as-if random is a sufficient but not
necessary condition for causal identification. Additionally, focusing
exclusively on the treatment assignment is not the most statistically
efficient way of dealing with the \emph{estimation} problem. And it is
misleading because the emphasis on \emph{treatment} assignment can make
us believe that we should focus exclusively on understanding how the
treatment came to be, without paying as much attention to understand the
determinants of the outcome; however, these are just two sides of the
same coin
\citep{LewbelIdentification2019a}\footnote{It is worth noting that the very assumptions about the so-called ``assignment mechanism'', like unconfoundedness, cannot be written down without mentioning the outcome itself. See, for example, \citet[p. 38]{ImbensCausal2015}.}.

\subsection{Structural Causal Model (SCM)}

\subsubsection{Structural Equations and DAGs}

The structural causal model (SCM) is built upon the symbiosis between
potential outcomes, non-parametric structural equations, and graphical
models. Instead of taking the individual potential outcomes as the
primitives from which we then aggregate to a target population, the SCM
is a population-first approach from where potential outcomes can be
derived. We start from our \emph{qualitative assumptions} about the data
generating process encoded as a set of structural equations (see the
first column in Table \ref{tab:dag_intro}), capturing functional or
mechanistic dependencies between variables; in other words, we ask, for
each variable, to what other variables in the system do they
\emph{causally} respond or ``listen to'' (to use a common metaphor).
These relationships are \emph{non-parametric} in the sense that we do
not restrict the dependency between variables to have any particular
form. We further assume that each variable have some idiosyncratic
component (everything else that affect their values, and that it is not
explicitly included in the model), that we denote as \(U_{V}\) for each
variable \(V\). Table \ref{tab:dag_intro} shows the simplest of such
systems encompassing three variables, which we can later use as the
building blocks for more complex models.

\begin{table}[h]
\begin{threeparttable}
\caption{Basic components of a Structural Causal Model (SCM)}
\label{tab:dag_intro}
\begin{tabular}{l c l c}
\toprule
Structural equations & Graphical model & Trunctated factorization & Conditional Independencies\\
\midrule
$A \mathrel{\vcenter{\baselineskip 0.5ex \lineskiplimit 0pt
                     \hbox{\scriptsize.}\hbox{\scriptsize.}}}%
                     =f_A(U_A)$ &  & $P(A,B,C)=$ &  $A \not\!\perp\!\!\!\perp C$\\
$B \mathrel{\vcenter{\baselineskip 0.5ex \lineskiplimit 0pt
                     \hbox{\scriptsize.}\hbox{\scriptsize.}}}%
                     =f_B(A, U_B)$  & $A \rightarrow B \rightarrow C$ & $P(A)P(B|A)P(C|B)$ & $A\perp\!\!\!\perp(C | B)$ \\
$C \mathrel{\vcenter{\baselineskip 0.5ex \lineskiplimit 0pt
                     \hbox{\scriptsize.}\hbox{\scriptsize.}}}%
                     =f_C(B, U_C)$ & (Chain)\\
\midrule
$A \mathrel{\vcenter{\baselineskip 0.5ex \lineskiplimit 0pt
                     \hbox{\scriptsize.}\hbox{\scriptsize.}}}%
                     =f_A(B, U_A)$ &  & $P(A,B,C)=$ & $A \not\!\perp\!\!\!\perp C$ \\
$B \mathrel{\vcenter{\baselineskip 0.5ex \lineskiplimit 0pt
                     \hbox{\scriptsize.}\hbox{\scriptsize.}}}%
                     =f_B(U_B)$  & $A \leftarrow B \rightarrow C$ & $P(B)P(A|B)P(C|B)$ & $A\perp\!\!\!\perp(C | B)$ \\
$C \mathrel{\vcenter{\baselineskip 0.5ex \lineskiplimit 0pt
                     \hbox{\scriptsize.}\hbox{\scriptsize.}}}%
                     =f_C(B, U_C)$ & (Fork)\\
\midrule
$A \mathrel{\vcenter{\baselineskip 0.5ex \lineskiplimit 0pt
                     \hbox{\scriptsize.}\hbox{\scriptsize.}}}%
                     =f_A(U_A)$ &  & $P(A,B,C)=$ & $A \perp\!\!\!\perp C$ \\
$B \mathrel{\vcenter{\baselineskip 0.5ex \lineskiplimit 0pt
                     \hbox{\scriptsize.}\hbox{\scriptsize.}}}%
                     =f_B(A,C,U_B)$  & $A \rightarrow B \leftarrow C$ & $P(A)P(C)P(B|C,A)$ & $A \not\!\perp\!\!\!\perp(C | B)$ \\
$C \mathrel{\vcenter{\baselineskip 0.5ex \lineskiplimit 0pt
                     \hbox{\scriptsize.}\hbox{\scriptsize.}}}%
                     =f_C(U_C)$ & (Collider)\\
\toprule
\end{tabular}
\end{threeparttable}
\end{table}

On the first column, we have the structural equations, defining in full
generality all endogenous variables as an unspecified function (\(f_V\))
of other variables and their respective error terms. In column 2, we see
how these structural equations can be represented using a Directed
Acyclic Graph (DAG), where each variable is a node in the graph, with
their causal dependency represented as links between nodes (for
simplicity, the error terms can be omitted from the graphs when assumed
independent from each other). In the language of graphs, the connection
between two variables is called a \emph{path}. More specifically, a
\emph{directed path} between two variables is a (set of) link(s) that
can bring us from one variable (node) to another, following the
direction of the arrows. Arrow direction represents the flow of causal
influence. We say that a node \(A\) is an ancestor of node \(B\) if it
takes causal precedence in a directed path. An immediate ancestor
(descendant) is called a parent (child). Since causality should respect
temporal order, the graph must be \emph{acyclic}, so the future cannot
cause the past. To be causally interpreted, a DAG should include all
common ancestors of every pair of variables included in the model, even
if unknown
\citep{ElwertGraphical2013, MorganCounterfactuals2014, PearlCausality2009}\footnote{DAGs can accomodate unknown common parents of any pair of variables by including an unobserved common parent $U$, usually encoded as a bidirected arc, in which case we say that the DAG is semi-markovian.}.

Let's turn for a second to a probabilistic description on the model.
Recall that, following the chain rule of probability, the joint
distribution of three random variables, i.e., \(P(A,B,C)\) can be
factorized as \(P(A)P(B|A)P(C|A,B)\) (or other equivalent forms).
However, by imposing some qualitative (causal) restrictions (i.e.,
indicating which variables listen to each other, excluding some possible
relations) we can afford a simpler factorization, and derive testable
implications from these factorizations. The third column in Table
\ref{tab:dag_intro} contains the simpler factorizations afforded by the
structural equations. The basic idea is that each variable's probability
depends only on its parents, and once parents are accounted for, the
rest of the variables do not carry additional
information\footnote{To be more precise, each variable is independent of non-parents and non-descendants, given the parents (\textit{local markov property})}.
If, for example, the joint probability factorizes according to the model
in the first row of Table \ref{tab:dag_intro}, we can deduce that \(A\)
should become independent of \(C\) within levels of \(B\). Similar
arguments can be made for the remaining models. The reader can check
that this is the case using the corresponding truncated factorizations.

One of the most powerful tools granted by the use of DAGs is that of
directed separation or \(d-\)\emph{separation}, which allow us to read
these expected statistical (in)dependencies directly from the graph (See
Table \ref{tab:dag_intro}, Column 4). First, think of edges as
transmitting (causal) information between nodes. We say that a directed
path between two nodes \(A\) and \(C\) is open (these variables are
\emph{d-connected}) if one can traverse such path from \(A\) to \(C\)
without colliding into an edge in the opposite direction. In Table
\ref{tab:dag_intro}, \(A\) and \(C\) are d-connected in the chain and
fork graphs, and d-separated in the collider graph (because we cannot
pass information through colliding edges). In other words, in the first
two cases we expect \(A\) and \(B\) to be statistically dependent, but
independent in the third case. Furthermore, we can turn open paths into
blocked paths (and vice versa) by conditioning, graphically represented
as boxing a node. In the chain and fork graphs, adjusting for \(B\)
d-separates \(A\) and \(C\) (blocking a previously open path), while in
the collider graph, adjusting for \(B\) render \(A\) and \(C\)
d-connected, and therefore statistically dependent (unblocking a
previously closed
path)\footnote{While the \textit{local markov property} mentioned earlier describes a subset of the independences implied by the graph, we call \textit{global markov property} to all independences implied by $d-$separation in the graph. This is a one-way implication: $d-$separation implies independence, but not (necessarily) the other way around.}.

\subsubsection{Estimand, Bias, and Identification in the SCM}

In the graphical framework, identification can be thought about as
isolating causal from non-causal paths. But, which paths are
\emph{causal} and which are not? With the help of our structural causal
model, we can imagine a minimal (``surgical'') intervention into our
system, directly \emph{assigning} or setting the value of the exposure
variable of interest. To see how, let's first look at the structural
equations. An intervention into our treatment \(D_i\) corresponds to
wiping out the function \(f_D(\cdot)\) that produces its observational
value, and replacing it with a constant, interventional value. For every
downstream variable, we would recursively replace the observational
\(D\) with the interventional \(d\) in their structural equation. This
would induce a new joint probability distribution for the variables in
the model, a so-called \emph{post-intervention} distribution, from which
we can compute our effects of interest.

Formally, this surgical intervention can be denoted using the
\emph{do-}operator. We can express our target intervention as
\(do(D=d)\), and our \emph{causal estimand} as, for example,
\(\mathbb{E}[Y_i \vert do(D=d)]\)\footnote{This is of course the simplest case, where the intervention corresponds to a particular value ($do[D=d]$). Nothing prevents us, however, from implementing more complex interventions with different assignment rules.}.
For the binary case, the \(ATE\) can be written using the \emph{do-}
notation as a contrast between two interventions:
\(\mathbb{E}[Y_i \vert do(D_i=1)] - \mathbb{E}[Y_i \vert do(D_i = 0)]\),
which is equivalent to the potential outcomes
\(\mathbb{E}[Y_{1i}] - \mathbb{E}[Y_{0i}]\). Under the structural causal
model, these quantities are not \emph{primitives} but the solution of
\(\mathbb{E}[Y_i|D_i=d]\) computed on the post-intervention
distribution\footnote{In other words, identification is about writing a post-intervention distribution in terms of a purely observational distribution, i.e., with values that can actually be computed from data, removing any \textit{do} along the way.}.

Of course, the ``easiest'' way to compute
\(\mathbb{E}[Y_i \vert do(D_i=d)]\) is by actually using
post-interventional data, like the one obtained from a randomized trial.
However, when we only have observational data, how can we approximate
the results we would obtain from an intervention without actually
intervening? How can we go from \emph{seeing} to \emph{doing}? To help
answering this question, we might benefit from using the DAG
representation. Recall that we want all \emph{causal paths} (i.e., those
starting from the exposure and reaching the outcome,
\(D \rightarrow \dots \rightarrow Y\)) to remain open, but all non
causal paths (also known as \emph{confounding} paths) to stay closed.
First, notice that, in the structural account, \emph{confounding} is not
anymore a property of variables in isolation, but a property of paths
related to a specific treatment-outcome pair.

Let's now turn to Table \ref{tab:dag_test}. Starting from an
observational graph \(G\), we can construct an \emph{interventional} DAG
\(G_{\overline{D}}\), where the arrows pointing into \(D\) were removed,
indicating that the value of \(D\) is set without attention to its
parents. In this graph, the paths \(d-\)connecting \(D\) and \(Y\) are
\emph{all causal}, so these are the paths we want to isolate. How do we
device a strategy to approximate the target graph? For this, we can
construct what is called a \emph{testing graph} \(G_{\underline{D}}\),
where all arrows \emph{coming out} from \(D\) (i.e., all causal paths)
are removed. The utility of this graph is double: on the one hand, since
all causal paths were removed from this graph, it allows to visualize
the \emph{sources of bias} in the observational data (i.e., all
remaining \(d-\)connections between \(D\) and \(Y\)); on the other hand,
using the testing DAG and the rules of \(d-\) separation, we can device
a way to isolate (\emph{identify}) the effect of \(D\) on \(Y\) by
blocking all paths between them that remain open in
\(G_{\underline{D}}\).

A few things are important to notice here. First, the non-causal paths
connecting \(D\) and \(Y\) in \(G_{\underline{D}}\) are known as
\emph{backdoor paths}, since they enter \(D\) ``from the backdoor''
(i.e., pointing into \(D\), in contrast to the causal paths ``going
out'' from \(D\) in \(G_{\overline{D}}\)). Second, the \emph{backdoor
criterion} \citep{PearlCausality2009, MorganCounterfactuals2014} taught
us that the effect of \(D\) on \(Y\) can be identified by conditioning
on a (set of) variable(s) \(X\) that (i) block every path between \(D\)
and \(Y\) in \(G_{\underline{D}}\), and (ii) do not block any path
between \(D\) and \(Y\) in \(G_{\overline{D}}\). Therefore, we have now
a new answer to the question of what to condition for to identify the
causal effect of \(D\) on \(Y\): the set \(X\) of variables that
satisfies the backdoor criterion.

\begin{table}[h!]
\begin{threeparttable}
\caption{Identification using graphical models}
\label{tab:dag_test}
\begin{tabular}{l c c c c c }
\toprule
& \multicolumn{2}{c}{A) $Y_d \perp\!\!\!\perp D | X$}  &&  \multicolumn{2}{c}{B) $Y_d \not\!\perp\!\!\!\perp D | X$}\\
\cmidrule{2-3} \cmidrule{5-6}
 & \textbf{Structural Equations} & \textbf{Graphical model}& & \textbf{Structural Equations} & \textbf{Graphical model} \\
\toprule
\multicolumn{6}{c}{(i) Observational graph $G$: the assumed data generating process}\\
&
\makecell[cl]{$X \mathrel{\vcenter{\baselineskip 0.5ex \lineskiplimit 0pt
                     \hbox{\scriptsize.}\hbox{\scriptsize.}}}%
                     =f_X(U_X)$ \\
$D \mathrel{\vcenter{\baselineskip 0.5ex \lineskiplimit 0pt
                     \hbox{\scriptsize.}\hbox{\scriptsize.}}}%
                     =f_D(X, U_D)$\\
$Y \mathrel{\vcenter{\baselineskip 0.5ex \lineskiplimit 0pt
                     \hbox{\scriptsize.}\hbox{\scriptsize.}}}%
                     =f_Y(D,X,U_Y)$}
 & \includegraphics[align=c, width=0.25\linewidth]{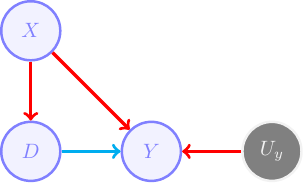} & & 
\makecell[cl]{$X \mathrel{\vcenter{\baselineskip 0.5ex \lineskiplimit 0pt
                     \hbox{\scriptsize.}\hbox{\scriptsize.}}}%
                     =f_X(U_X)$ \\
$L \mathrel{\vcenter{\baselineskip 0.5ex \lineskiplimit 0pt
                     \hbox{\scriptsize.}\hbox{\scriptsize.}}}%
                     =f_L(U_L)$\\
$D \mathrel{\vcenter{\baselineskip 0.5ex \lineskiplimit 0pt
                     \hbox{\scriptsize.}\hbox{\scriptsize.}}}%
                     =f_D(X, L, U_D)$\\
$Y \mathrel{\vcenter{\baselineskip 0.5ex \lineskiplimit 0pt
                     \hbox{\scriptsize.}\hbox{\scriptsize.}}}%
                     =f_Y(D,X,L, U_Y)$}&  \includegraphics[align=c, width=0.25\linewidth]{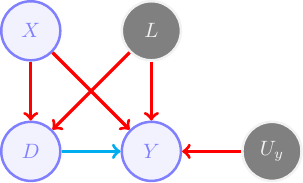} \\
\midrule
\multicolumn{6}{c}{(ii) Interventional (target) graph $G_{\overline{D}}$: remove all variables into $D$}\\
&
\makecell[cl]{$X \mathrel{\vcenter{\baselineskip 0.5ex \lineskiplimit 0pt
                     \hbox{\scriptsize.}\hbox{\scriptsize.}}}%
                     =f_X(U_X)$ \\
$D \mathrel{\vcenter{\baselineskip 0.5ex \lineskiplimit 0pt
                     \hbox{\scriptsize.}\hbox{\scriptsize.}}}%
                     =d$\\
$Y \mathrel{\vcenter{\baselineskip 0.5ex \lineskiplimit 0pt
                     \hbox{\scriptsize.}\hbox{\scriptsize.}}}%
                     =f_Y(d,X,U_Y)$} & \includegraphics[align=c, width=0.25\linewidth]{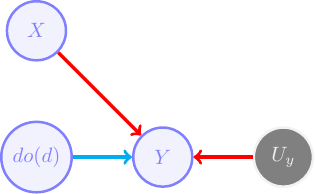} & & 
\makecell[cl]{$X \mathrel{\vcenter{\baselineskip 0.5ex \lineskiplimit 0pt
                     \hbox{\scriptsize.}\hbox{\scriptsize.}}}%
                     =f_X(U_X)$ \\
$L \mathrel{\vcenter{\baselineskip 0.5ex \lineskiplimit 0pt
                     \hbox{\scriptsize.}\hbox{\scriptsize.}}}%
                     =f_L(U_L)$\\
$D \mathrel{\vcenter{\baselineskip 0.5ex \lineskiplimit 0pt
                     \hbox{\scriptsize.}\hbox{\scriptsize.}}}%
                     =d$\\
$Y \mathrel{\vcenter{\baselineskip 0.5ex \lineskiplimit 0pt
                     \hbox{\scriptsize.}\hbox{\scriptsize.}}}%
                     =f_Y(d,X,L, U_Y)$} &  \includegraphics[align=c, width=0.25\linewidth]{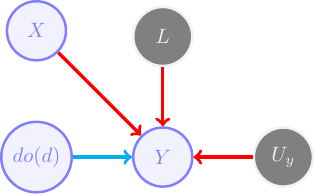} \\
\midrule
\multicolumn{6}{c}{(iii) Testing graph $G_{\underline{D}}$: remove all variables from $D$}\\
&
\makecell[cl]{$X \mathrel{\vcenter{\baselineskip 0.5ex \lineskiplimit 0pt
                     \hbox{\scriptsize.}\hbox{\scriptsize.}}}%
                     =f_X(U_X)$ \\
$D \mathrel{\vcenter{\baselineskip 0.5ex \lineskiplimit 0pt
                     \hbox{\scriptsize.}\hbox{\scriptsize.}}}%
                     =f_D(X, U_D)$\\
$Y \mathrel{\vcenter{\baselineskip 0.5ex \lineskiplimit 0pt
                     \hbox{\scriptsize.}\hbox{\scriptsize.}}}%
                     =f_Y(X,U_Y)$} & \includegraphics[align=c, width=0.25\linewidth]{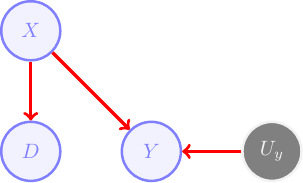} & & 
\makecell[cl]{$X \mathrel{\vcenter{\baselineskip 0.5ex \lineskiplimit 0pt
                     \hbox{\scriptsize.}\hbox{\scriptsize.}}}%
                     =f_X(U_X)$ \\
$L \mathrel{\vcenter{\baselineskip 0.5ex \lineskiplimit 0pt
                     \hbox{\scriptsize.}\hbox{\scriptsize.}}}%
                     =f_L(U_L)$\\
$D \mathrel{\vcenter{\baselineskip 0.5ex \lineskiplimit 0pt
                     \hbox{\scriptsize.}\hbox{\scriptsize.}}}%
                     =f_D(X, L, U_D)$\\
$Y \mathrel{\vcenter{\baselineskip 0.5ex \lineskiplimit 0pt
                     \hbox{\scriptsize.}\hbox{\scriptsize.}}}%
                     =f_Y(X,L, U_Y)$} &  \includegraphics[align=c, width=0.25\linewidth]{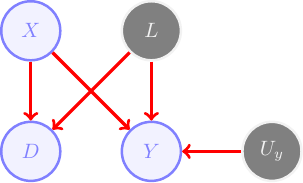} \\
\midrule
\multicolumn{6}{c}{(iv) Conditioning graph $G_{X=x}$: remove all variables from $X$}\\
&
\makecell[cl]{$X \mathrel{\vcenter{\baselineskip 0.5ex \lineskiplimit 0pt
                     \hbox{\scriptsize.}\hbox{\scriptsize.}}}%
                     =f_X(U_X)$ \\
$D \mathrel{\vcenter{\baselineskip 0.5ex \lineskiplimit 0pt
                     \hbox{\scriptsize.}\hbox{\scriptsize.}}}%
                     =f_D(X=x, U_D)$\\
$Y \mathrel{\vcenter{\baselineskip 0.5ex \lineskiplimit 0pt
                     \hbox{\scriptsize.}\hbox{\scriptsize.}}}%
                     =f_Y(D,X=x,U_Y)$}
 & \includegraphics[align=c, width=0.25\linewidth]{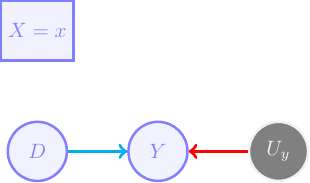} & & 
\makecell[cl]{$X \mathrel{\vcenter{\baselineskip 0.5ex \lineskiplimit 0pt
                     \hbox{\scriptsize.}\hbox{\scriptsize.}}}%
                     =f_X(U_X)$ \\
$L \mathrel{\vcenter{\baselineskip 0.5ex \lineskiplimit 0pt
                     \hbox{\scriptsize.}\hbox{\scriptsize.}}}%
                     =f_L(U_L)$\\
$D \mathrel{\vcenter{\baselineskip 0.5ex \lineskiplimit 0pt
                     \hbox{\scriptsize.}\hbox{\scriptsize.}}}%
                     =f_D(X=x, L, U_D)$\\
$Y \mathrel{\vcenter{\baselineskip 0.5ex \lineskiplimit 0pt
                     \hbox{\scriptsize.}\hbox{\scriptsize.}}}%
                     =f_Y(D,X=x,L, U_Y)$}&  \includegraphics[align=c, width=0.25\linewidth]{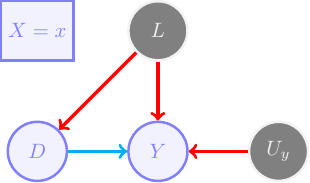} \\
\midrule
\multicolumn{6}{c}{(v) Evaluate using graphical criteria: can we $d-$separate $D$ and $Y$ in $G_{\underline{D}}$?}\\
&
\multicolumn{2}{c}{Yes, conditioning on $X$} &&
\multicolumn{2}{c}{No, because $L$ is unobserved}\\
\multicolumn{6}{c}{Alternatively, evaluate using potential outcomes: are the arguments in $f_Y(\cdot) \perp\!\!\!\perp D \vert X=x$?}\\
&
\multicolumn{2}{c}{$D \perp\!\!\!\perp x$} && 
\multicolumn{2}{c}{$D \perp\!\!\!\perp x$}  \\
&
\multicolumn{2}{c}{$D \perp\!\!\!\perp U_Y$} && 
\multicolumn{2}{c}{$D \perp\!\!\!\perp U_Y$}\\
&&&& \multicolumn{2}{c}{$D \not\!\perp\!\!\!\perp L$}\\
\bottomrule
\end{tabular}
\end{threeparttable}
\end{table}

We are finally in condition to connect back to the potential outcomes
framework. As noted above, the PO framework is a unit-first approach,
with the individual potential outcomes \(Y_{di}\) being \emph{defined}
as the what-if outcomes a unit \(i\) would have if exposed to condition
\(D_i=d\), everything else equal. Naturally, different interventions
would produce different potential outcomes for different units. The SCM,
on the other hand, is a population-first approach, where the outcome
variable \(Y\) is \emph{defined} as a function of their parents in
\(f_Y(\cdot)\), and idiosyncratic variation between units is subsumed in
\(U_Y\) following some unknown distribution. Therefore, if we set
\(u_Y = i\) (i.e., random variation represented by a unit indicator),
the individual potential outcomes can be equivalently written as
\(Y(do(d), u_Y) = Y(d, u_Y) = Y_d(u_Y) = Y_{di}\). Following this logic,
it can be shown that the backdoor criterion is the \emph{graphical
equivalent} to the conditional exchangeability assumption. To see how,
let's look at row (v) on Table \ref{tab:dag_test}. Since the potential
outcome \(Y_d\) is nothing else than the solution to
\(f_Y(d,\dots, u_Y)\), we can evaluate if \(Y_d\) is independent from
the treatment \(D\) by checking if the arguments in \(f_Y(\cdot)\) are
(conditionally) \(d-\)separated from \(D\) in the graph \(G\).

For concreteness, let's go back to the question of the returns of
education. For the sake of the example, assume that receiving College
education is determined by your parents' socioeconomic status (\(X\)).
In a second, more realistic model, assume that College going is also
determined by unobserved or latent factors (\(L\), such as academic
motivation and family expectations). Additionally, we might suspect that
the factors affecting one's education have, simultaneously, an impact on
one's income. Table \ref{tab:dag_test} (first row) depicts these two
scenarios (with and without the latent factor), with the causal effect
of interest represented by the path \(D \rightarrow Y\), and the
confounding (non-causal) paths by \(D \leftarrow X \rightarrow Y\) and
\(D \leftarrow L \rightarrow Y\). Examining the graph, we can
immediately see the threat for identification: in the observational DAG
\(G\), \(D\) and \(Y\) are d-connected by more than one path, and
therefore the observational association between them is a mixture of
causal and non-causal relations. From the same graph, we can assess if
the effect of interest is identified. In the left panel, the answer is
positive: adjusting for \(X\) blocks all non-causal paths. However, in
the right panel, the answer is negative, due to the existence of the
unobserved variable \(L\).

\subsubsection{Heuristics: Deriving testable implications}

If the heuristics for researchers using the potential outcomes framework
is thinking by analogy to a randomized experiment, following the logic
of the ``assignment mechanism'', in the case of graphical models the
proposed heuristics is taking advantage of \(d-\)separation to read off
testable implications from the assumed model. For example, under the
logic of the assignment mechanism, it is common to provide evidence in
favor of an adjustment-based strategy by showing ``balance'' between
treatment groups and across (an ideally large number of) covariates, as
this is what one would expect to observe under a randomized assignment.
Under the SCM, on the other hand, this expected independence is but a
special case of all the testable implications that can be derived by
inspecting what variables are \(d-\)separated (or connected) in a given
DAG.

For example, a particularly powerful version of this happens when,
according to the model, there exist more than one admissible adjustment
set, i.e., more than one group of variables that satisfy the backdoor
criterion. Instead of conditioning by the maximum number of variables
possible (\emph{all covariates at the same time}), one could separately
adjust for the sets that are minimally sufficient; if they provide
consistent answers to our query, we have stronger evidence than what we
would obtain by including everything in a single
estimation\footnote{See, for example, the discussion of models 1-6 in \cite{CinelliCrash2022}}.
Recall that, as previously discussed, confounding is a property of
paths, not variables: there is no need to adjust for all pre-treatment
variables, only to block all non-causal paths.

Another important consequence of this approach, is realizing that
confounding should always be assessed based on a specific
treatment-outcome pair, and not the treatment alone. For example, in
observational studies, it is all too common to see the same propensity
score model used to estimate the causal effect of a given treatment on
different outcomes, under the assumption that, since all that matters is
the assignment mechanism, once it is correctly modeled the specific
outcome becomes irrelevant. However, as can be easily shown with the
help of graphical models, a set of variables that is sufficient to
remove confounding for one outcome might not be so for a different one.
Something similar can be said of instrumental variable settings, where
assessing the plausibility of an exclusion restriction critically
depends on our understanding of the outcome determinants, and not only
on the exogeneity of the instrument.

One last example of situations where the use of graphical models can
result helpful is when conducting falsification/placebo tests and
sensitivity analysis. In the case of falsification/placebo tests, they
are frequently motivated in a rather informal way. Researchers select
variables that are assumed not to be affected by the treatment, but
without necessarily providing a formal reasoning behind such assumption.
Graphical models, on the other hand, provide a way of systematically
deriving which variables should, or should not, be (conditionally)
independent from each other, following the rules of \(d-\)separation. In
the case of sensitivity analysis, researchers provide estimates of the
maximum amount of ``hidden confounding'' (omitted variable bias) that
would be admissible before changing a study's conclusions; however,
without reference to what those ``hidden confounders'' could be, and
their structural relation with the variables already included in the
model, it is very difficult to have a productive discussion about the
plausibility of such scenario, at least beyond some vague skepticism.

\subsection{Summary and preliminary balance}

The main point of the previous section was to show that potential
outcomes and graphical models need not be, at least \emph{in theory},
opposed to each other. Quite the contrary, complementing the PO
definitions with graphical identification criteria offers the tools to
assess how plausible some identification assumptions are, based on our
qualitative understanding of the data generating process: we just need
to write down our assumptions, and we can derive their consequences for
identification\footnote{The epidemiologist Miguel Hernán has a well-known virtual course on using graphical models for causal inference, that invites to ``Draw your assumptions before your conclussions''. But for some critics of graphical models this seems to be an insane advice, as I will discuss in the following section.}.
There are, however, several differences on how \emph{in practice} each
community tend to approach empirical research. While disciplines with a
preference for using potential outcomes have focused on understanding
how a particular treatment comes to be, with an increasing preference
for quasi-experimental designs and growing skepticism for purely
observational studies, users of graphical models do not shy away from
messy settings where there is no source of clearly ``exogenous''
variation. This has opened a gulf on how the credibility of empirical
research is assessed in each community. In what follows, I will argue
that it is possible to benefit from the use of graphical models while
remaining agnostic with respect to the settings in which they are being
applied, and without committing to any \emph{ex ante} understanding of
the data generating process.

\section{The case for research templates and the controversy over DAGs}

If, as I have argued, DAGs are so useful to transparently encode our
assumptions and derive their implications, then why haven't graphical
models been more widely adopted in empirical research, facing instead
some passionate resistance? In this section, I attempt to provide a
partial answer to this question, related to the notion that, to be
credible, causal analysis should follow certain templates and that
deviation from those templates necessarily decreases the credibility of
causal inferences. I also describe the relation between the template
model and the use of quasi-experiments in empirical research. In its
more extreme form, the quasi-experimental ideal involves a complete
rejection of graphical models as unnecessary or even dangerous; however,
it can also be accompanied by a recognition of their \emph{pedagogical}
utility. Then, I discuss a prevalent description of how DAGs should be
developed (a process that I refer as the ``maximal DAG'' approach),
showing how it feeds into the skepticism regarding graphical models. I
end up hinting at a more flexible and ``minimal'' approach to DAGs, that
I believe it can help reconcile the way causal inferences are assessed
in practice and across disciplines.

\subsection{The case for research templates}

\begin{quote}
\emph{Partly as a result of the focus on empirical examples, the
econometrics literature has developed a small number of canonical
settings where researchers view the specific causal models and
associated statistical methods as well-established and understood. These
causal models correspond to what is nowadays often referred to as}
identification strategies \hfill --- Imbens (2020)
\end{quote}

As highlighted by the opening quote, \emph{identification strategies}
are well-known sets of assumptions that are sufficient to justify the
causal interpretation of certain estimators. With an original push
coming from economics, but increasingly adopted in other social
sciences, the idea that credible causal research should follow these
pre-defined templates is seductive: we know what quantities they recover
(their \emph{estimand}), we know under which qualitative conditions they
work (their \emph{identifying assumptions}), and we know what
statistical analysis to pair them with (the appropriate
\emph{estimators}). An enthusiastic defense of this approach can be
found in the essay on the relevance of potential outcomes and graphical
models for empirical economics by the nobelist Guido Imbens. There,
underscoring the simplicity of the template model, Imbens recalls that
``many of the current popular identification strategies focus on models
with relatively few (sets of) variables, where identification questions
have been worked out once and for all''
\citep[p. 1131]{ImbensPotential2020}.

In this ``working out'', the potential outcomes framework has played a
prominent role. On the contrary, Imbens does not seem at all convinced
that adding graphical models to a researcher's toolkit would be as
useful. He provides two main reasons for this skepticism: on the one
hand, the preference for simplicity makes the value added by DAGs almost
negligible given the favored identification strategies; on the other
hand, while DAGs seems powerful to derive identification results in
complex settings, those settings are not well suited for credible
identification. To put it in a sentence: credible models of empirical
research are too simple for DAGs to be useful, and models where DAGs can
prove useful are too complex to be credible. Let's attend to each
argument in more detail.

\subsubsection{Too simple to be useful}

One example of the (lack of) utility of DAGs when dealing with common
identification strategies is the case of unconfoundedness (a.k.a.,
conditional ignorability or selection on observables). Referring to the
illustrative diagram in Figure \ref{fig:Imbens_SOO},
\citet[p. 1164]{ImbensPotential2020} argues that, in cases like this,
using graphical models do not provide much insight beyond what potential
outcomes offer, because:

\begin{quote}
\begin{quote}
``this setting, where the critical assumption is ignorability or
unconfoundedness, is so common and well studied, that merely referring
to its label is probably sufficient for researchers to understand what
is being assumed. Adding a DAG, or for that matter adding a proof that
the average causal effect is identified in that setting, is superfluous
because researchers are familiar with the setting and its
implications''.
\end{quote}
\end{quote}

Two things are worth pointing out from this quote. First, when confining
ourselves to the displayed template, it is certainly hard to disagree
with Imbens' judgment: this scenario is so idealized that identification
is immediately apparent, either by looking at the graphical model or the
corresponding potential outcomes. Therefore, if for DAGs to be useful
they should make a difference in \emph{deriving} the identifying
assumptions of an idealized case in a pre-defined set of canonical
identification strategies, one may concede they are likely to fail the
test (at least in the eye of an
econometrician)\footnote{Some recognizable exceptions where graphical models have made a difference in deriving identifying conditions are the front-door criterion \citep{PearlCausal1995}, work on time-varying settings \citep{RobinsMarginal2000a, PearlProbabilistic1995}, and mediation analysis \citep{PearlDirect2001}.}.
In this context, Imbens argue, using graphs or not is all a ``matter of
taste'' \citep[p. 1138]{ImbensPotential2020}.

\begin{center}
\includegraphics{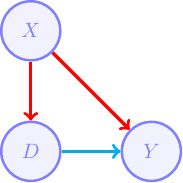}
\captionof{figure}{DAG for unconfoundedness, where $D$ is the treatment, $Y$ the outcome, and $X$ a vector of \textit{confounders}.}
\label{fig:Imbens_SOO}
\end{center}

But this bring us immediately to the second point, closely related: what
are we making assumptions \emph{about} when conducting empirical
research? Are we making assumptions about a template, in which case
Imbens' view is spot on, or are we making assumptions about a setting
that can be arbitrarily far from the template? In fact, the point of
disagreement seems to be partially due to differences in judgement on
how likely (or how acceptable) is to be far removed from the textbook
case. Even more, one may ask, who is the one making those assumptions
and assessing their plausibility? The methodologist deriving new
identification results, the researcher claiming to follow a given
template, the reviewer assessing the plausibility of the invoked
assumptions, or the reader trying to understand what is in fact being
assumed? The relevance of graphical model for empirical research hinges
critically on the answer to these questions.

This is noted by Imbens himself, when indicating that ``DAGs are often
clear and accessible ways to expressing visually some, though not
necessarily all, of the key assumptions. It may provide the reader, even
more than the researcher, with an easy way to interpret and assess the
overall model''
\citep[p. 1155]{ImbensPotential2020}\footnote{And he immediately continues: ``I am less convinced that the formal identification results are easier to derive in the DAG framework for the type of problems commonly studied in economics''.}.
I believe this is the crux of the matter: while for researchers (let
alone econometricians) their causal framework of preference might as
well be ``a matter of taste'`, for the scientific community as a whole
it is irreplaceable to have a transparent \textit{lingua franca} to
avoid each researchers' solipsism. In this regard, graphical models are
much more difficult to disregard.

\subsubsection{Too complex to be credible}

One might ask if, when dealing with more complex situations, graphical
models can be put to a better use. In fact, one of the benefits of
graphical models is that, even with multiple nodes and many edges, the
basic rules of \(d-\)separation keep things simple, and identification
can even be automated given a certain
DAG\footnote{Several softwares and statistical packages do this already. See, for example, \href{www.dagitty.net}{www.dagitty.net} and \href{www.causalfusion.net}{www.causalfusion.net}.}.
But this bring us back to Imbens' second argument: the cases where the
machinery of DAGs can be useful, are at the same time the cases where
causal claims are less credible. In these situations, the argument goes,
the so-far trivial DAG transmutes into its opposite, going from
superfluous tool to ``dangerous master'', one that ``can easily lull the
researcher into a false sense of confidence''
\citep{ImbensDiscussion1995}. Or, in the words of
\citet{RosenbaumDiscussion1995}, the very fact that causal diagrams can
handle complex settings and interconnected assumptions is undesirable
``because it is much more difficult to ensure that the assumptions are
warranted''\footnote{Going back to Imbens when discussing the problem of covariate selection and M-bias, he recognizes that, without DAGs, there is no clear answer. However, he remind us, ``In order to use these methods, one needs to carefully consider every absent link, and in a setting with as many variables as there are in figure 13, panel B [5 variables], that is a daunting task'' \citep[p. 1166]{ImbensPotential2020}. But, is it not the whole point that, even in template-adjacent cases, we are still required to make numerous assumptions that the DAG itself is not introducing, just bringing to the forefront?}.

Recent discussions about what credible assumptions in empirical research
amount to have stretched the causal inference community even further
away. While Imbens discusses the importance of sticking to known
templates (and including selection on observables in his list), some go
a step further and argue that only the subset of templates --those that
exploit some source of ``exogenous variation'' in the exposure of
interest-- deserves serious consideration. Take, for example, the
influential \emph{Mostly Harmless Econometrics}
\citep{AngristMostly2009}, where the authors emphasize the need of
looking for ``well-controlled comparisons and/or natural
quasi-experiments'' in the absence of randomization. Other forceful
defenses of the quasi-experimental ideal can be found in
\citet{ImbensBetter2010a} and \citet{DoleacEvidenceBased2019}. According
to this view, the role of the researcher is to find real-world
situations that mirror the quasi-experimental template (at most
requiring some minor fixes), thus avoiding the temptation of going
astray with adjustments and conditioning strategies. The spirit is
well-captured in the words of the economist Jason Abaluck (cited in
\cite{ImbensPotential2020}):

\begin{quote}
\begin{quote}
``No one should ever write down a 100 variable DAG and do inference
based on that. That would be an insane approach because the analysis
would be totally impenetrable. Develop a research design where the 100
variable DAG trivially reduces to a familiar problem (e.g.~IV!).''
\end{quote}
\end{quote}

At this point, a few things are worth discussing in more detail. Notice,
first, how the phrasing ``\emph{trivially} reduces'' is doing a lot of
the heavy lifting. But even conceding that reduction to a template is
trivial, at least in quasi-experiments (I will later argue that this is
\emph{not} the case in practice), one may still wonder if the only two
options we have as researchers are either picking cases with three or
four variables, or cases with a hundred of them. Building on examples, I
will argue that empirical research is better described as in-between
these two extremes, and that graphical models are a useful tool to
transparently argue about our attempts to ``reduce to a template.''
However, before that, let's look at one use of DAGs that have gained a
more universal acceptance: their role as teaching tools.

\subsubsection{Graphical scaffoldings}

\begin{quote}
\emph{The fact that DAGs are not useful for everything is no argument
against them. All theory tools have limitations. I have yet to see a
better tool than DAGs for teaching the foundations of and obstacles to
causal inference. And general tools like DAGs have added value in
abstracting away from specific details and teaching us general
principles} \hfill --- McElreath (2020)
\end{quote}

After being advised against their dangers, one if left wondering if DAGs
have any fruitful use. In this regard, there is some convergence on a
``weak'' case for graphical models: the increasing consensus that causal
diagrams are enormously helpful teaching tools. While
\emph{identification strategies} has been worked out ``once and for
all'' in the history of human knowledge, these venerable methodological
insights need to be inscribed in the minds of every new generation of
researchers. In this sense, it is recognized that DAGs are wonderful
intuition-building devices, transparently capturing some critical
assumptions in causal inference, thus facilitating the learning of
research designs. Just like an scaffolding, however, graphs can be
discarded once the solid building is finished, i.e., when more seasoned
researchers have tamed their causal intuitions (one might be tempted to
repeat ``once and for all'').

It is in this capacity that DAGs appear frequently in textbooks and
educational articles. For example, in the best selling \emph{Causal
Inference: The Mixtape}, \citet{CunninghamCausal2021} argues that
graphical models are, among other things, ``very helpful for
communicating research designs and estimators if for no other reason
than pictures speak a thousand words''. Similar uses of DAGs can be
found in \citet{BrumbackFundamentals2021, LiuQuantifying2021}, and
\citet{WestreichEpidemiology2019}. For each identification strategy
being introduced, there is an accompanying graphical model that
supposedly ``corresponds'' to it. While pedagogically useful, this
approach is not far removed from the template ideal, treating graphical
models more as accessory illustrations rather than as tools for causal
reasoning in actual empirical research.

\subsection{Where do DAGs come from? The maximal DAG approach}

\begin{quote}
\emph{At this point you may be wondering where the DAG comes from. It's
an excellent question. It may be the question. A DAG is supposed to be a
theoretical representation of the state-of-the-art knowledge about the
phenomena you're studying. It's what an expert would say is the thing
itself, and that expertise comes from a variety of sources} \hfill ---
Cunningham (2021)
\end{quote}

\begin{quote}
\emph{DAGs are a particularly useful tool for causal inference because,
when one has correctly specified the hypothetical DAG to encode prior
knowledge of the data generating structure, a set of mathematical-based
rules can be applied to determine whether and how the causal effect of
interest can be identified. However, the DAG framework is entirely
silent on what the causal model should look like, and epidemiologists
often struggle to tie their methodologic training in DAGs to specific,
real-world settings} \hfill --- Matthay and Glymour (2020)
\end{quote}

It is the elephant in the room: the little guidance that exists on how
to construct and refine graphical causal models aimed to empirical
researchers\footnote{Here I am not referring to the different and more ambitious task of directly discovering the causal structure or ``true DAG'' underlying observational data. For an up to date introduction to that research agenda, see \cite{PetersElements2017}.}.
The question is, then, how can we, as researchers, use graphical models
to encode \emph{what we already know} (or what we are able to suppose)
in order to decide if we can learn \emph{what we still don't know}?
Should we actually start from scratch and write down all possible
variables and causal relations relevant to our research questions in the
abstract, or is this an attempt destined to fail? Should we rather stick
to already proven templates and research designs, simple enough to not
require graphical aid? Is there anything in between?

I suspect that DAG-skepticism comes from what can be called the
``maximal DAG'' approach. For advocates of this deductive, top-down
process, what is required is to develop a DAG in full generality,
\emph{context-less}, starting purely from theory and available evidence.
Putting it succinctly, according to this view, the DAG construction
process can be summarized in three steps: 1) define a research question,
identifying its associated pair of exposure and outcome, 2) build a
bivariate DAG, connecting the exposure and outcome through the
hypothesized causal effect, 3) expand, including all sources of
association between the pair that one can think of, across all contexts
and applications. See, for example, this process described in
\citet{Laubachbiologist2021}\footnote{The authors describe a DAG as a ``flowchart that maps out th causal and temporal relationship between X and Y, along with additional variables that might affect the $X \rightarrow Y$ association'', which are listed as (i) confounders, (ii) precision covariates, (iii) mediators, (iv) effect modifiers, and (v) colliders.},
and
\citet{WestreichEpidemiology2019}\footnote{``For example, if we wish to know the effect of $X$ on $Y$, we would start by drawing a node labeled $X$, a node labeled $Y$, and a single-headed arrow from $X$ to $Y$ indicating the possibility of a causal effect of $X$ on $Y$ [...] From this base, we create other nodes for other variables we believe to be relevant to the $X \rightarrow Y$ relationship.''}.
A more complete and systematic approach can be found in
\citet{PetersenCausal2014}, including the discussion on how to
incorporate uncertainty about the true data-generating process.

This deductive process can be hard to swallow. For one thing, it entails
the risk of growing without limits, as suggested by the technical
requirement that for a DAG to be ``causal'' it has to include all common
parents for every pair of variables included in the graph
\citep{ElwertGraphical2013, PearlCausality2009}\footnote{This can certainly be relaxed using semi-markovian graphs, where missing common parents are replaced by bidirected arrows, and whose identification rules are known.}.
It also assumes that the researcher is in position to describe the main
components and causal relations in the relevant data generating process,
without omitting any relevant aspects. This produces a paradoxical
result: we may end up with a monstrous graph (a hundred variables!)
that, at the same time, risks forgetting precisely \emph{the} confounder
that hinders identification.

However, one may argue that this DAG-induced anxiety is an integral part
of conducting research. We know that everything has to do with
everything; we also know that, without abstracting, learning is not
possible. This is why some authors have highlighted the need of trimming
the diagram once we have incorporated all variables and relations that
we \emph{suspect} may be relevant \citep{Huntington-KleinEffect}. The
critical step is then arguing why certain nodes and arrows can be
omitted from the diagram, based on previous research, theoretical
assumptions, and institutional
knowledge.\footnote{``The willingness to make a stand and say 'yes, we really can omit that' --that's what you need to answer your research question! It is truly necessary. Anybody trying to answer a research question who won't do that is still making those assumptions implicitly. They just aren't making clear what those assumptions are.'' \citep{Huntington-KleinEffect}}
In other words, based on a
model\footnote{This does not mean that the influence of the variables omitted from the model has to be \textit{exactly} zero. One way of thinking about this is in terms of the order of magnitude of the influence of included versus excluded variables, paired with calibrated sensitivity analysis as in \citet{CinelliMaking2020}. Or, following \citet{ShaliziAdvanced2021}, ``what we really have to assume is that the relationships between the causes omitted from the DAG and those included is so intricate and convoluted that it might as well be noise''.}.

An important caveat is necessary here. Being fully algorithmic once the
DAG is defined, identification do not requires that we remove nodes and
arrows for \emph{computational convenience}. The simple rules of
d-separation can provide us with testable implications from the diagram.
And, no matter how big and complex the graph, an identification
algorithm can return the answer to our query, if there is any allowed by
the DAG. This is an important benefit of using causal diagrams, because
we can focus on encoding our qualitative assumptions without fearing
that the model becomes intractable. On the other hand, this very feature
is seen with suspicion by some: a powerful distraction, allowing
researcher to entertain the possibility of distinguish association from
causation in severely lost cases.

In summary, growing and trimming a DAG is just a formal and transparent
way of sharing our understanding about the data generating process,
which informs our analyses. In this context, scientific criticism and
theory refinement is what occurs in the struggle to erase and redraw
nodes and arrows, to incorporate, omit, or break variables into smaller
units. As we will see, one of the most overlooked benefits of using DAGs
is that they facilitate this conversation: not only providing
transparency on the researchers' side (what are we assuming?), but also
help to formalize our criticisms about research credibility (where are
we disagreeing on?). In that sense, DAGs are wonderful tools for
\emph{conceptual replication}.

\subsection{DAGs for research diagnostics}

Certainly, as with any powerful tool, one needs a proper amount of
discipline to not get hurt by careless manipulation; but this is in no
way an indictment against the tool itself. There is, of course, the
danger of simply omitting things from the DAG to secure identification,
making convenience-based instead of knowledge-based assumptions, as
warned by \citet{PetersenCausal2014}. Even in that case, however, what
is being assumed-away is transparently encoded in the DAG, which is
offered to the scrutiny of critics. The risk of becoming too confident
about our own model, as feared by many
\citep{AronowReview2020, ImbensDiscussion1995, RosenbaumDiscussion1995},
may always be confronted by productive scientific criticism. If
anything, having a visual representation of our assumptions should help
this process. In other words, what graphical models afford for us is
simply this: being able to focus on \emph{substantive} assumptions,
rather than arguing on the grounds of computational convenience.

In this sense, using graphical models to capture the minimal features of
research templates falls short in exploiting the expressive capabilities
of causal diagrams. When sticking to the template model, we are at the
same time asking \emph{too much} and \emph{too little} from DAGs:
\emph{too much} since, in many settings, they do not provide an
exhaustive list of the required identification assumptions; but
\emph{too little}, since we are restricting ourselves to highly stylized
templates that only tenuously resemble what researchers actually do,
instead of using graphical models tailored to capture the specificity of
each setting.

In this section, I have collected several examples from the literature
showing how DAGs, when used flexibly, turn apparent problems that
otherwise might go unnoticed, while indicating ways to increase the
credibility of causal inferences if possible. In fact, contrary to the
overconfidence narrative, many scholars have used graphical models to
highlight the shortcomings of common research designs, indicating why
certain quantities of interest are not identified, against popular
wisdom. Table \ref{tab:dag_examples} contains examples from a variety of
settings, including research on social networks, multigenerational
mobility, the returns of education, and police brutality. Let's look in
more detail at some of these examples.

\begin{sidewaystable}[ph!]
\centering
\caption{Some examples of DAGs used as templates of common research problems}
\label{tab:dag_examples}
\begin{threeparttable}
\begin{tabular}{c c c c c}
\toprule

\textbf{Research} & \textbf{Graphical} & & & \\
\textbf{Question} & \textbf{Model} & \textbf{Diagnosis} & \textbf{Proposal} & \textbf{References}\\
\midrule

 \makecell[cl]{Can we identify the effect \\ of grandparents' social \\position on a person's own \\position, net of the \\effect of their parents?} & 
 \includegraphics[align=c, width=0.2\textwidth]{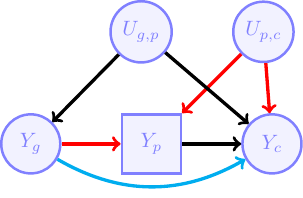} &
 \makecell[cl]{Collider bias estimating \\the direct effect of \\grandparents' ($Y_g$) \\on children' status ($Y_c$)\\ conditioning on \\parental status ($Y_p$)} &
 \makecell[cl]{Consider alternative \\explanations of the \\observed associations\\ and alternative designs} &
 \makecell[cl]{Breen (2018) discussing \\ Erola abd Niusui (2007) \\Chan and Boliver (2013) \\ Hertel and Groh-\\Samberg (2014) \\Ziefle (2016)}\\

 \makecell[cl]{} & 
 \includegraphics[align=c, width=0.2\textwidth]{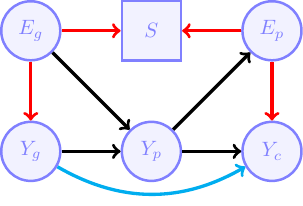} &
 \makecell[cl]{Collider bias estimating \\the direct effect of \\grandparents' ($Y_g$) \\on children' status ($Y_c$) \\due to conditioning \\on joint survival ($S$)} &
 \makecell[cl]{Adjusting for age\\ at first birth \\($E_g$ and $E_p$), \\ sensitivity analysis \\ and IV designs} &
 \makecell[cl]{}\\
\midrule

 \makecell[cl]{Can we use police \\administrative records \\to estimate racial \\discrimination in\\ policing?} & 
 \includegraphics[align=c, width=0.2\textwidth]{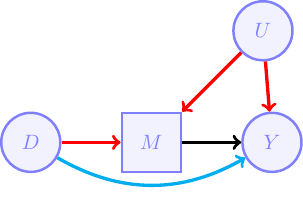} &
 \makecell[cl]{Target quantity unclear, \\ collider bias estimating \\direct effect of race\\even within selected \\sample of stops} &
 \makecell[cl]{The authors propose\\ a bias-correction method, \\a bounding procedure, \\and a design that is \\not affected by the \\highlighted problems.} &
 \makecell[cl]{Knox et al. (2020) \\ discussing Antonovics \\ and Knight (2009) \\Fryer (2018, 2019) \\ Johnson (2019) \\ Ridgeway (2006)}\\
\midrule

 \makecell[cl]{Can we distinguish \\between homophily \\(selection) \\and contagion \\(peer effects) \\ in social networks?} & 
 \includegraphics[align=c, width=0.25\textwidth]{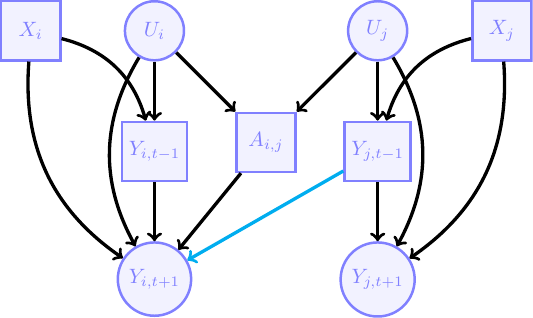} &
 \makecell[cl]{Conditioning on a collider \\ ($A_{i,j}$, i.e., the existence \\of a tie between \\individuals $i$ and $j$) \\impedes disentangling \\both effects, \\and current methods do \\ not work without (strong) \\further assumptions} &
 \makecell[cl]{Exhaustively adjust for \\homophily determinants, \\ use community detection \\in social networks, \\estimate bounds} &
 \makecell[cl]{Shalizi and Thomas (2011) \\ discussing Aral et al. (2019) \\ Anagnostopoulos et al. (2008) \\ Bakshy et al. (2009) \\Bramoullé et al. (2009) \\ Christakis and Fowler \\(2007, 2008) \\ Yang et al. (2007)}\\
\midrule

 \makecell[cl]{Is intergenerational mobility \\higher among college \\graduates than non \\graduates? (i.e. is a\\ college degree \\the great equalizer?)} & 
 \includegraphics[align=c, width=0.2\textwidth]{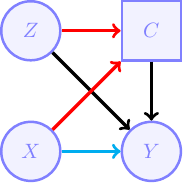} &
 \makecell[cl]{Current estimates \\suffer from collider bias \\(conditioning on $C$, \\college completion), \\thus distorting estimates\\ of intergenerational mobility \\ ($X \rightarrow Y$)} &
 \makecell[cl]{Residual balancing, \\a method to break the link\\ between $Z$ (jointly \\determinants of college $C$\\ and income $Y$)\\ and college completion ($C$) \\before conditioning on $C$, \\for unbiased estimation \\of $X \rightarrow Y$} &
 \makecell[cl]{Zhou (2019) discussing\\ Hout (1984, 1988) \\ Breen (2010) \\ Chetty et al. (2017)\\ Pfeffer and Hertel (2015) \\ Torche (2011)}\\
\bottomrule
\end{tabular}
\end{threeparttable}
\end{sidewaystable}

In the first row, we can see how \citet{BreenMethodological2018} used
graphical models to highlight some serious issues in the study of
multigenerational mobility. In particular, several attempts have been
made to identify and estimate the effects of grandparents on children's
status attainment, net of (``above and beyond'') the effects of the
parental generation. The problem with this is encoded in the first DAG
in Table \ref{tab:dag_examples}: assuming, as it is plausible to do,
that there are unobserved factors (\(U_{p,c}\)) affecting the social
attainment of both the parental (\(Y_p\)) and children (\(Y_c\))
generations, then conditioning on the status of the parental generation
(\(Y_p\)), a collider variable, distorts the association of the
grandparent's and children's generations (\(Y_g \rightarrow Y_c\)) by
opening a non-causal path between them
(\(Y_g \rightarrow Y_p \leftarrow U_{p,c} \rightarrow Y_p\)). The
problem is so serious that there is little hope to directly estimate the
effect of interest in such design. The proposal to improve the analysis
is, therefore, to either look at alternative explanations of the
association of interest (an indirect approach), or simply finding a
different design that is not subject to the same
problems\footnote{Notice, additionally, that this problem is \textit{structurally equivalent} to the problem presented in the second example, regarding racial discrimination in policing. Using graphical models helps revealing hidden similarities in the issues faced by different subfields, and can therefore facilitate methodological cross-pollination.}.

The second diagram in the first row shows a different problem in the
study of multigenerational mobility. One may be interested in the
mechanisms of the influence of grandparents on grandchildren; one of
such hypothesized mechanism is the length of the exposure of children to
their grandparents, as measured by cohabitation or joint survival
(\(S\)). However, as shown in the diagram, the length of joint survival
is a function of the parent age at birth in both generation (\(E_g\),
\(E_p\)), which itself depends on, and affect subsequent socioeconomic
status (\(Y_g\),\(Y_c\)). Therefore, conditioning on joint survival
opens a biasing path between the grandparent and children generation.
Once again, this subtle issue is easily overlooked without the help of
graphical models.

The DAGs in Table \ref{tab:dag_examples} are also, in a sense,
\emph{templates}, but they are so in a very different sense than the
previously discussed identification strategies. They operate on a
different \emph{level of abstraction} than both over-simplified
identification strategies and over-complicated top-down graphical
models. In contrast to research templates encoding totally abstract
identification strategies, these graphical models speak directly to a
substantive domain, to a collection of empirical findings. And, contrary
to building DAGs top-down, starting purely from (usually context-less)
theory, these graphical models were constructed by making explicit the
assumptions buried under a given set of empirical applications. This
process is inductive: from a collection of examples (empirical studies),
a common pattern is derived, a \emph{structural similarity} is
identified. Once translated into a DAG, we can see what these studies
have in common, including their shared problems. From there, possible
solutions can be found to be tested in new studies, contributing to
improve research credibility for an entire subfield (or even across
them).

This use of graphical models is likely the closest version of what I
call a ``minimal DAG'' approach, showing us the way to get the most our
of DAGs. From here, we only need to make explicit one last step:
formalizing the missing link between graphical models such as those
included in Table \ref{tab:dag_examples}, and the actual examples and
applications form the empirical literature they are based on. In other
words, we need better guidelines on how to use DAGs in a more granular,
study-specific level of analysis, to assess the credibility of actually
existing research. This is what I am to describe in the next section.

\section{Moving beyond templates of empirical research}

\begin{quote}
\emph{The answer essentially is an indirect one: if you tell me how the
world works (by giving me the full causal graph), I can tell you the
answers. Whether this is satisfactory really revolves around how much
the researcher is willing to assume about how the world works}
\hfill --- Imbens (2020)
\end{quote}

\begin{quote}
\emph{Graphs offer a disciplined framework for expressing causal
assumptions for entire systems of causal relationships {[}\ldots{]} This
advantage is especially apparent when the analyst must consider the many
causal pathways that may be present in the real applications of
observational research} \hfill --- Morgan and Winship (2015)
\end{quote}

In this section, I argue that graphical models are most useful, not when
used to represent idealized features of perfectly implemented designs,
and then ascribing a certain study to its corresponding category, but
when used to describe a study's particular setting. I flesh out the
proposal of a ``minimal DAG'' approach, introducing the idea of using
graphical models to encode the \emph{evidence generating process} (as
opposed to the \emph{data generating process}), and proposing a workflow
to build DAGs bottom-up. I demonstrate the approach through applied
examples, reconstructing the assumptions implicit in some
quasi-experimental studies, to show that the distinction with
observational studies is not clear-cut and that research credibility is
better understood as a matter of degree rather than a hierarchy of
designs, highlighting the limits of using template labels as a shortcut
to assess the credibility of empirical findings.

\subsection{Towards a minimal DAG approach}

There seems to be an underlying tension between interpreting graphical
models as a scientific program, versus understanding graphical models as
scientific tools. As a program, so the critics perceive, using graphical
models implies endorsing an overly optimistic view of how much do we
know about the world. At the very minimum, this would mean assuming
researchers are able write down the ``true DAG'', the full causal graph,
at least in a given application domain. At its maximum, it would imply
that we can even \emph{discover} the underlying structure of reality by
purely algorithmic means. On the other hand, using graphical models as a
tool for causal reasoning, does \emph{not require} endorsing such views:
we can still use graphical models as ``a disciplined framework for
expressing causal assumptions'' while remaining agnostic about our
knowledge of the full data generating process. And this framework can
prove useful even if (or especially if) one favors quasi-experimental
studies as the standard of credible empirical evidence.

Indeed, we can only go so far by sticking to idealized templates, and
soon enough researchers feel compelled to claim all sorts of
``conditional'' strategies, even in quasi-experimental settings: an
instrument might only be as-good-as-random after adjusting for certain
covariates, potentially including alternative mechanisms that affect the
outcome; there could be a true regression discontinuity only for a given
subpopulation, or we may want to extrapolate far from the cutoff; the
timing of a policy can be seen as exogenous only in certain geographic
areas; and so on. In all such cases, explicit or implicitly, one needs
to deal with assumptions about the structure of the entire system of
causal relations under consideration, even if that is just a clipping of
the full graph.

Trained in the template model, researchers may find themselves lacking
the tools to rigorously adapt their preferred identification strategy to
their actual setting. We are then left in the disjunctive of either
abstracting away from our particular application and proceed as-if we
were in the template case, thus risking erroneous inferences, or
avoiding altogether situations in which the templates cannot be applied,
missing valuable opportunities for conducting empirical analyses. Facing
this situation, it is all too common for researchers to fall back into
``hopeless flirtation with regression'' (quoting Judea Pearl), while
still claiming their place in the \emph{credibility ladder}
corresponding to the invoked template. Are we condemned to this state of
affairs, or can graphical models help us to bring some clarity into this
identification gray zone?

\subsection{Modeling the Evidence Generating Process}

One way of conceptualizing the role that graphical models can play in
empirical research, that may help us going beyond the dichotomy between
totally general and hardly credible DAGs, and more credible but highly
idealized (and, sometimes, oddly specific) research templates, would be
to distinguish between different levels of abstraction in the research
process. In particular, I would like to introduce the distinction
between the \emph{data generating process} and the \emph{evidence
generating process}. The proposed conceptual model is depicted in Figure
\ref{fig:DAG_egp}, Panel (A).

For \emph{data generating process} (DGP) we usually understand the
entire system of causal relationships and sources of randomness that
give rise to the observable data; in other words, the DGP is what
actually happens in the world. Of course, knowing the \emph{true} DGP
would be sufficient to identify and compute any counterfactual we may be
interested in, and using DAGs or potential outcomes for that effect
would certainly be a matter of taste. By a wide margin, however, this is
not generally the position social scientists are in.

While researchers usually do not claim to understand the full DGP of the
problem they are studying, they do claim to understand the particular
setting of their empirical analyses. For example, in a
quasi-experimental study, researchers would claim to understand in great
detail the process that produced a certain exposure, to the point of
judging that the resulting treatment assignment was as-good-as-random.
Generally, however, that very same judgment indicates the need of
adjusting for some background confounding variables, or alternative
mechanisms that may bias the effect of interest if left unaccounted for.
In other words, the \emph{evidence generating process} (EGP) is the
particular setting in which we claim to be able to identify a given
causal effect. While the DGP is highly general, the EGP is a much more
specific clipping of reality.

Notice that, when researchers claim to identify a causal effect using a
given identification strategy, the invoked template must be contained in
the EGP at hand, which is itself contained in the more general DGP
(Figure \ref{fig:DAG_egp}, Panel (A)). But jumping from the general DGP
to the template is far from trivial; that reduction needs to be argued
and assumptions need to be made explicit. Using graphical models for
causal reasoning, we provide ourselves with the tools to make such
assumptions as explicit as possible, and open to scrutiny from other
researchers. Practically, this does not require to build DAGs top-down,
from theory alone, starting from an \emph{ex-ante} understanding of the
DGP. Instead, one can go ahead as usual with a given quasi-experiment,
and only then use graphical models to write down our assumptions in an
inductive fashion: what assumptions are \emph{implied} in the way we
analyzed a given quasi-experiment? In other words, we can write the
model starting from the empirical design, rather than from theory.

\begin{center}
\begin{minipage}{0.65\linewidth}
\includegraphics[width=\linewidth]{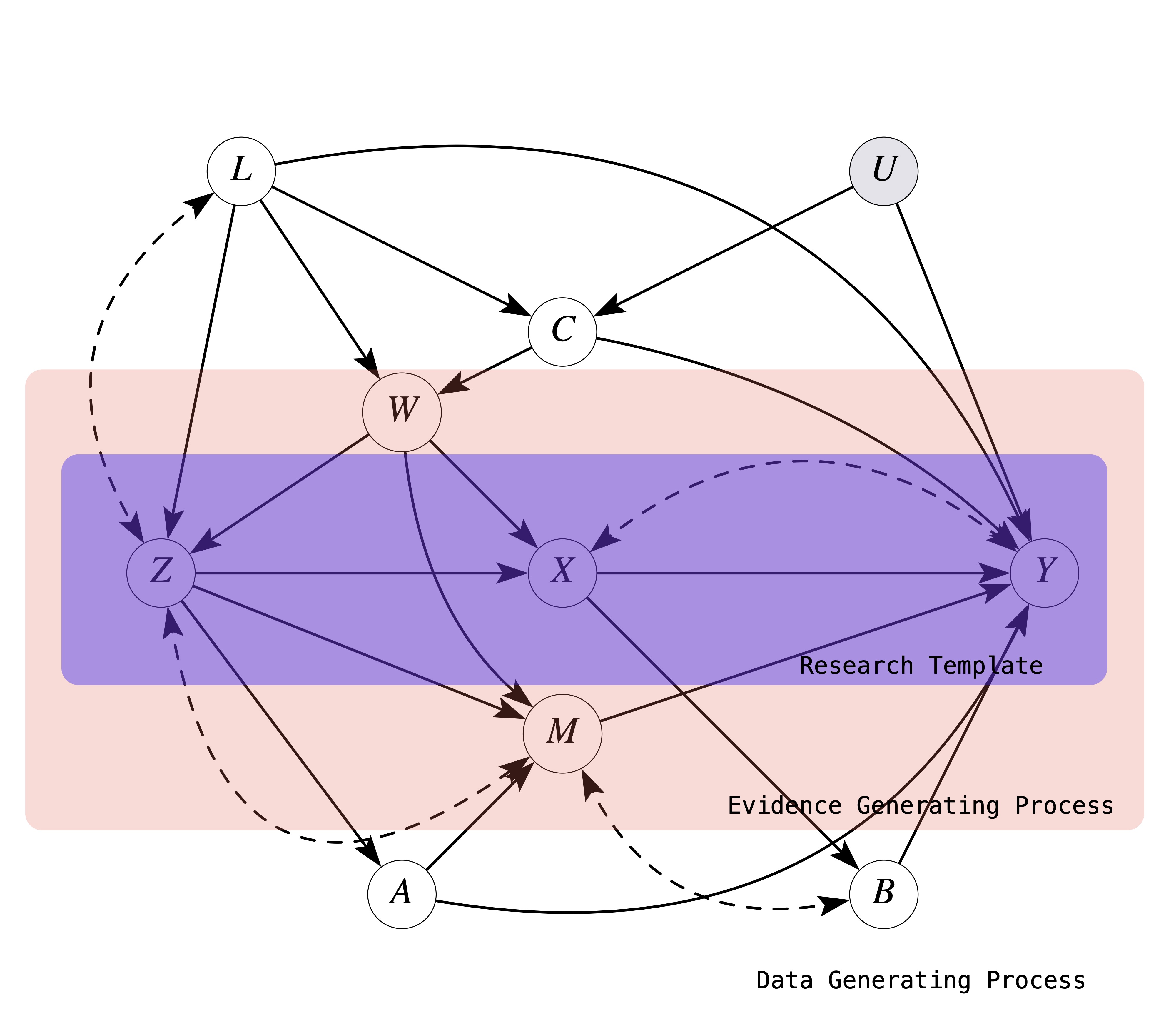}
\captionof*{figure}{\textbf{(A)} Different levels of abstraction in research.}
\end{minipage}%
\hfill
\begin{minipage}{0.65\linewidth}
\includegraphics[width=\linewidth]{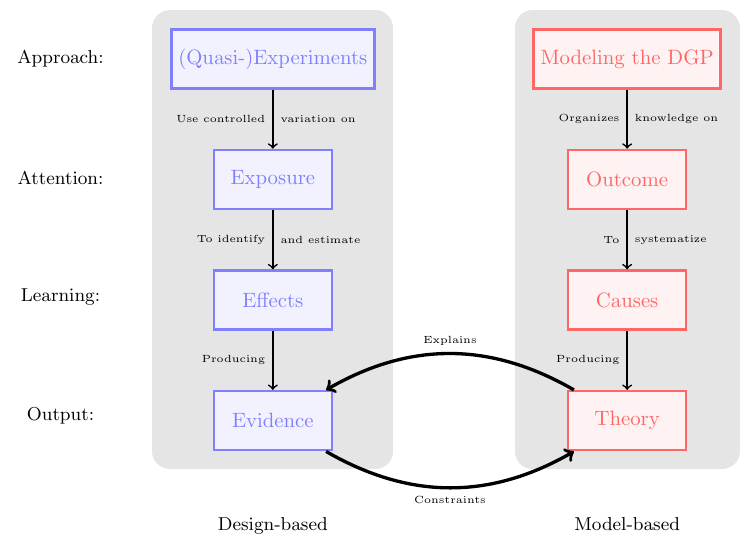}
\captionof*{figure}{\textbf{(B)} Dynamic relation between the design-based and model-based stages in research practice.}
\end{minipage}
\captionof{figure}{Conceptual model: How to use DAGs for empirical research.}
\label{fig:DAG_egp}
\end{center}

\subsection{The Theory-Evidence-Theory' circle}

Another look at the dynamics of the research process can be helpful to
understand the point (Figure \ref{fig:DAG_egp}, Panel (B)).
Quasi-experimental research has been described as a design-based
approach: using controlled or well understood variation on a given
exposure or treatment variable, one can identify and estimate its
effects, resulting in a collection of evidence claims. For example: a
reform changing the minimum school leaving age (well-understood source
of variation) induces an extra year of education (the treatment) for a
given cohort (the exposed population). We can then identify the impact
or effect of that extra year of schooling on labor market outcomes,
producing evidence of school impacts. Different reforms, affecting
different cohorts at different ages would, of course, create different
results.

If stopping here, we are still in the realm of impact evaluation. In the
process of scientific research, however, we would want to use that
evidence to impose \emph{constraints} on our theories of education:
let's say, schools have so and so effects on income, and any good theory
of schools, or income, should account for that. Or the other way around:
evidence from experimental or quasi-experimental studies may show that
aspects previously hypothesized as important were in fact negligible, so
they can be safely omitted from the model.

We may as well start this iterative process from somewhere else, for
example, organizing our knowledge about a given outcome (income
distribution), systematizing their known causes (which may come from
previous impact evaluations), so that we construct a theory
\emph{explaining} or making sense of the evidence. In our example, we
may interpret the increased income of more formally educated people as
coming from their human capital, or a return to their credentials, among
other possible stories. The point is that, lacking a model of the
evidence generating process, we cannot move from doing impact evaluation
to doing social science.

Some researchers may focus in producing evidence while others may focus
in creating explanations that conform to that evidence. However, the
distinction is highly schematic, and in reality both sides of the
scientific practice are highly intertwined. To discard alternative
explanations and, therefore, interpreting a given association as
\emph{evidence} of a causal effect, one needs to have at least a working
understanding of the process giving rise to the outcome of interest.
That understanding may be weak, incomplete and partial, a mere
scaffolding; it is nonetheless there, and using graphical models to make
it explicit is just our due service to research transparency and
scientific dialogue. Because, quoting Philip Haile, ``without a model
there is often only hand waving''.

\subsection{Building DAGs from the bottom-up: a proposed workflow}

One could describe the proposed approach as an \emph{agnostic} use of
causal diagrams to assess the credibility of empirical results \emph{as
they appear} in practice. What I mean by agnostic in particular is that
the causal model, and its corresponding graphical representation, should
not be (or at least does not need to be) the starting point. Graphical
models constructed bottom-up or inductively, starting from the
assumptions authors are \emph{actually} willing to make, can be
enormously useful to assess the plausibility of the invoked assumptions,
and using them do not require us to commit to a full generative model
\emph{ex ante}. In other words, graphical models can be used as a
contrast agent to visualize if we are able to stand our own assumptions
about the evidence generating process once they are brought to the
forefront.

This is not, in principle, opposed to the use of quasi-experiments. It
is only to say that, to fulfill their promise of increasing the
credibility of causal inference, quasi-experiments should be treated as
\emph{observational studies}. Their character as ``highly plausible''
ones is precisely what is at stake and it should be critically examined
rather than ritually accepted. I argue that this approach to graphical
models reveals that the distinction between purely observational and
quasi-experimental is not clear-cut, with different shades of grey in
between that are better illuminated by explicitly encoding the
assumptions in a transparent way, taking full advantage of the
expressive capabilities of graphs. This way, DAGs become what they are
meant to be: neither trivial nor dangerous, but useful tools for
transparent scientific communication.

Following a few simple steps, summarized in Table
\ref{tab:step_by_step}, can help us to get the most out of graphical
models when assessing the credibility of empirical research, including
in quasi-experimental settings. They are not all necessary in a given
application, nor they provide an exhaustive list of relevant question to
ask. But these steps offer an starting point for a more systematic
conversation when assessing empirical claims about causal relations.

Let's begin with the middle column, that contains the suggested
procedure: which steps to take in order to build the DAG starting from
the implicit assumptions. We can also see that, for each step, there are
a few questions that immediately arise, showing us the implications of
our assumptions along the way. On the left column, I highlight the
benefit of different steps, in terms of increasing the transparency of
our assumptions, deriving testable implications of our models, and
generating more substantive and productive scientific engagement.
Finally, the right column describe the source from which each step
derives: either from assertions internal to the study at hand, from
formal graphical rules, or from external knowledge in the form of models
or expert opinion.

\begin{table}[ph!]
\begin{threeparttable}
\caption{Assessing research credibility using graphical models: A step-by-step guide}
\label{tab:step_by_step}
\begin{singlespace}
\begin{tabular}{l | l | l}
\toprule
Benefit & Procedure & Source\\
\midrule
Transparency & \makecell[cl]{\textbf{1. Construct the causal diagram inductively}\\
When planning or reading a particular study, one should ask: \\
\quad \textit{What variables are being included in the analysis?}\\
\quad \textit{What relationships between variables are being assumed?} \\
\\
\textbf{2. If necessary, build a set of possible models}\\
If it is not possible to arrive to a single model
implied by the study,\\ build a group of them:\\
\quad \textit{Can the models be reduced to the same structure?} Simplify!\\
\quad \textit{Do the models contradict each other?} Keep them all\\
\\
\textbf{3. Evaluate the identification argument} \\
Is the quantity of interest identified under the authors' own assumptions?\\
\quad \textit{Yes, in all models}: Continue down\\
\quad \textit{No, at least in some models}: Requires clarification} & \makecell[cc]{Internal to the study +\\ graphical rules}\\
\midrule
Testability & \makecell[cl]{
\textbf{4. Systematically derive testable implications}\\ 
For each model, one should ask ``If the model is true, then'':\\
\quad \textit{Which variables should be associated?}\\
\quad \textit{Which variables should be independent/balanced?}\\
\\
\textbf{5. Test compatibility with observed data}\\
\quad \textit{Is the data compatible with the testable implications?}\\ 
\quad \textit{How much can we relax the assumptions before the results change?}\\
\\
\textbf{6. Assess the internal credibility of the assumptions}\\
\quad \textit{Is there any missing relationship between the variables already included?}\\ 
\quad \textit{Does this change the conclusions?}}& \makecell[cc]{Implied model +\\ graphical rules +\\expert opinion}\\
\midrule
Generativity & \makecell[cl]{\textbf{7. Assess the external credibility of the assumptions }\\
\quad \textit{Is there any missing variable that should have been included?}\\
\quad \textit{Does this change the conclusions?}\\
\\
\textbf{8. If necessary, build an alternative diagram}\\ 
Taking into account the results from the model criticism exercise\\
\quad \textit{Add necessary but previously omitted variables}\\
\quad \textit{Add necessary but previously omitted relationships}\\
\\
\textbf{9. Update and repeat}\\
With the new model at hand, the process starts over.} & \makecell[cc]{External evidence +\\ formal models}\\\\
\toprule
\end{tabular}
\end{singlespace}
\end{threeparttable}
\end{table}

From the list above, one might need to implement only a subset of the
steps depending on the circumstances. For example, a researcher
conducting an observational study (including, of course,
quasi-experiments), should routinely go from steps 1 to 5, and ideally
from 1 to 7. Journal reviews would enormously benefit by explicitly
engaging in steps 4 to 7, checking the internal consistency of the
researchers' working models, and their compatibility with the provided
evidence. This has the potential to create more productive exchanges
between authors and reviewers, where not only the authors' assumptions
but also the reviewers' objections are formalized and made explicit.
Steps 8 and 9, on the other hand, usually go beyond what a single study
can achieve, but are crucial components of a research agenda and the
development of the scientific literature as a whole.

\subsection{Some worked examples}

In this section, I illustrate the minimal approach to using graphical
models I have advocated for: building them from the bottom-up, starting
from actual research designs, to more transparently encode and assess
the assumptions made by the authors, even if implicitly. I will
demonstrate the approach analyzing two empirical pieces recently
published on top sociology journals, \emph{American Sociological Review}
and \emph{American Journal of Sociology}. The purpose of this exercise
is not to discuss in detail the findings of each article; rather, it is
to show how many unnoticed assumptions are better appreciated when
encoded in a graphical form, and that actual empirical research (even
under the rubric of quasi-experimental designs) can significantly depart
from the ideal template cases.

\subsubsection{Community and the Crime Decline (Sharkey et al. 2017)}

In ``Community and the Crime Decline: The Causal Effect of Local
Nonprofits on Violent Crime'' \citep{SharkeyCommunity2017}, the authors
aim to identify the effect of ``informal sources of social control
arising from residents and organization internal to communities'' on the
significant crime decline that the US experienced during the 1990-2010
period. Traditional explanations of this decline usually direct their
attention to factors extrinsic to communities, in particular policing
and harsher legal prosecution against crime. In contrast, the purpose of
the authors is trying to estimate if there is any effect of local
organizations, a factor intrinsic to the communities, on the crime
decline.

The treatment of interest are \emph{nonprofits focused on reducing
violence and building stronger communities}, while the outcome is
\emph{violent crime} (violent crime, homicide, and property crime). The
difficulty of estimating such an effect arises, according to the
authors, from the endogeneity of nonprofit formation. Therefore, in
addition to OLS and Fixed Effect models, they develop an instrumental
variable strategy to address the endogeneity
problem\footnote{``Because community organizations are formed at least partly in response to social problems like violence, it is not possible to rely on cross-sectional data and standard analytic methods to identify the effect of nonprofit formation on crime rates. To account for the endogeneity of nonprofit formation we use variation in the prevalence of nonprofits across cities and over time in a fixed-effects framework and adapt an instrumental variable (IV) strategy to identify the causal effect of nonprofits on crime'' \citep[p. 1215]{SharkeyCommunity2017}}.
The authors highlight the importance of their instrumental variable
approach because previous research ``have not dealt with endogeneity of
nonprofit formation'' and theirs would be ``the first study that
provides a plausible causal estimate of the impact of such organizations
on crime'' \citep[p. 1219]{SharkeyCommunity2017}. Therefore, in this
section I will focus on the IV setting, showing 1) how a real
application of an IV strategy departs from the template DAG discussed in
the previous section; 2) how graphical models can help us to assess the
credibility of such claims, and see where and why this strategy can
fail.

\textbf{A non-standard instrumental variable setting}

The authors describe their approach as relying on variation in the
formation of nonprofits that is created by factors not directly related
to the crime rate. Inspired by the strategy followed in
\cite{LevittUsing2002}, where firefighter hiring is used as a
(conditional) instrument for police hiring, Sharkey and colleagues
propose the use of nonprofit organizations for the arts, medical
research, and enviromental protections as an instrument for community
nonprofit. Why do the authors expect that both types of nonprofits,
community and non-community oriented, would be associated? In their own
words, the ``core assumption is that changes in the prevalence of arts,
medical, and evironmental nonprofits have no direct effect on crime and
violence but are associated with changes in the prevalence of nonprofits
designed to address violence and rebuild communities ---which, for
simplicity, we refer to as ``community'\,' nonprofits--- through common
mechanisms of funding availability.''
\citep[p. 1215]{SharkeyCommunity2017}.

To summarize, the basic identification assumptions are: 1) The number of
other nonprofits is related to the number of community nonprofits; 2)
after accounting for a set of control variables, there is no association
between other nonprofits and crime other than the association between
different types of
nonprofits\footnote{An additional assumption, not discussed here, is \textit{monotonicity}. Basically, this implies that there is no $City \times Year$ where an increase in other nonprofits would decrease the number of community nonprofit. Given budget constraints, this is unlikely to be the case.}.
Figure \ref{fig:Sharkey} provides the causal graph corresponding to the
authors' own description of their setting.

\bigskip

\begin{center}
\includegraphics{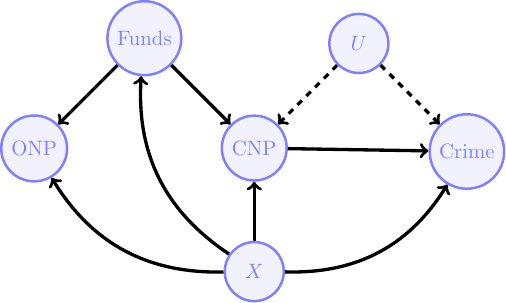}
\captionof{figure}{Instrumental Variable setting from Sharkey et al. (2017). The variables are: availability of funding (Funds), community nonprofits (CNP), other nonprofits (ONP), registered violent crime (Crime), and unobserved common factors between CNP and Crime ($U$). Control variables in $X$ are: population density, ethnic composition, educational composition, sex by age composition, immigration percentage, unemployment, and occupational composition.}
\label{fig:Sharkey}
\end{center}

\bigskip

A first thing to note is that, contrary to the template DAG, the
surrogate instrument (\(ONP\)) \emph{does not} cause the treatment
(\(CNP\)) nor it is exogenous (they share a common cause). It is likely
that many researchers would fail to recognize this as complying with the
IV conditions if only looking for examples that fit the simplest
template. However, this is still a valid instrument (conditional on
covariates in \(X\)) \emph{according to the graph}: 1) \(ONP\) is
\(d-\)connected to \(CNP\) through the path
\(ONP \leftarrow \text{Funds} \rightarrow CNP\), 2) \(ONP\) is
\(d-\)separated to \(Crime\) in the graph \(G_{\overline{CNP}}\), in
other words, in te graph where we \(do(CNP=d)\).

\textbf{Is the (surrogate) instrument confounded?}

As highlighted above, the instrumental variable conditions are meet
\emph{according to the graph}, so the next step is to consider how
credible is the graph itself. One first aspect to attend in this regard
is the adjustment set the authors consider sufficient for their IV
strategy to be valid. They do not actually provide a detailed rationale
for its inclusion, and the variables are listed as ``controls'\,'
without further justification \citep[p. 1219]{SharkeyCommunity2017}.
This is an important point because the validity of the design rests
critically on this step. Are they including those variables because they
are concerned with violations of the exogeneity assumptions? Or are they
trying to block possible direct effect of their instrument on the
outcome? The reader is leave without a clue about this.

To be more specific, this step is nothing else than an
\emph{observational study} embedded within the quasi-experimental design
highlighted in the front-end of the paper. The (surrogate) instrument is
recognized to be initially confounded, and therefore the authors are
making the (implicit) claim that the variables included in \(X\) can
make the instrument assignment conditionally independent of the
potential outcomes. In other words, this is the same assumption
underlying regression, matching, or weighting models: the level of
\emph{other nonprofits} a neighborhood has would be assigned based on
observed characteristics of the neighborhood. However, this get lost as
of secondary importance, and no theoretical justification nor
sensitivity analysis are provided to evaluate this assumption. It is not
hard to imagine that such an assumption would be subject to a much more
severe scrutiny have the authors declared as their main purpose to
identify the effect of \emph{other nonprofits}, or \emph{funding
availability}, on Crime. But, somehow, the reader is invited to believe
that the instrumental variable design is superior for causal
identification than ``observational'' alternatives, despite one of its
cornerstones is being left mostly unattended.

What type of variables would made us concerned about exogeneity
violations? Are they included in \(X\) already? Figure
\ref{fig:Sharkey_EX} provides two examples of possible violations.
\textbf{Panel (A)} represents the case of a factor \(L_1\) that affects
both the amount of resources available for nonprofits, and the crime
rate. One can imagine, for example, that the political orientation of
local government would affect the resource allocation for nonprofits,
while at the same time would affect the funding and strategy of
policing, one of the major determinants of crime rates as recognized by
the
authors\footnote{It is important to note that some models include $State$ and $Year$ fixed effects. However, this is not enough to adjust for the political seasonality, which would only be captured by an $State \times Year$ specification. The authors also provide a robustness check by including \textit{size of the police force}. However, this check suffers from two limitations: the sample size is substantially reduced due to data availability, and the size of the police could not be the best proxy for policing strategies and funding for their operation.}.
\textbf{Panel (B)} represent a different scenario. Here \(L_{t-1}\)
denote the unobserved (previous) level of community organization and
engagement, possibly informal and therefore not captured by registered
nonprofits. The arrows pointing from \(L_{t-1}\) to \(ONP\), \(Funds\),
and \(CNP\) represent the fact the communities themselves can bring
resources to their neighborhoods and manage their own nonprofits, and
the arrow \(L_{t-1} \rightarrow \text{Crime}\) captures the direct
effect of community involvement in
Crime\footnote{Once again, it is important to note that this concern is consistent with the authors' own literature review, where they state ``Organization within a community are embedded within larger networks of public and private agencies and organizations extending across a city's neighborhoods and beyong the city limits. These extra-local networks connect communities to external sources of influence, resources, and political power, all of which strengthen the capacity to achieve common goals and values (...) Communities with stronger internal and external ties, higher levels of social cohesion, and greater informal social control are more likely to be able to regultate activity in public spaces and control the threat of violence.'' \citep[p. 1218]{SharkeyCommunity2017}.}.

\begin{center}
\begin{minipage}{0.48\linewidth}
\includegraphics[width=\linewidth]{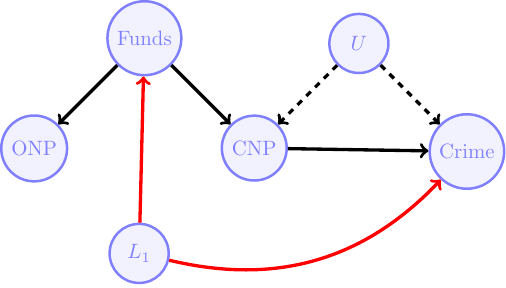}
\captionof*{figure}{\textbf{(A)} The (unobserved) instrument shares a common cause with the outcome. For example, political orientation of the local government could affect funds available for community organization and policing programs simultaneously.}
\end{minipage}%
\hfill
\begin{minipage}{0.48\linewidth}
\includegraphics[width=\linewidth]{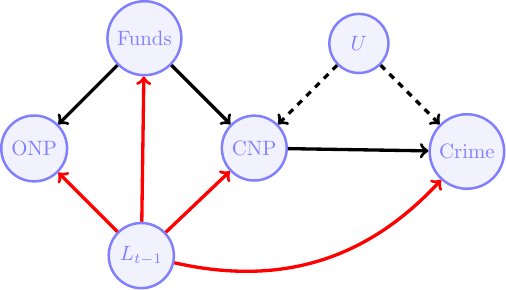}
\captionof*{figure}{\textbf{(B)} The instrument, the surrogate instrument and the treatment share a common cause with the outcome. For example, previous levels of community involvement could affect the availability of funds and have a direct effect on crime.}
\end{minipage}
\captionof{figure}{Possible violations of the exogeneity (unconfoundedness) assumption.}
\label{fig:Sharkey_EX}
\end{center}

\textbf{Is the effect of the instrument fully mediated by the
treatment?}

The second identification assumption in the instrumental variable
setting is called \emph{exclusion restriction}, which basically means
that the entire effect of the instrument on the outcome should be
mediated by the treatment. In this case, this implies that any influence
of \emph{other nonprofits} on Crime should be due to the causal effect
of \emph{community nonprofits} on Crime. Is this a plausible assumption
in our setting? Figure \ref{fig:Sharkey_ER} represents a possible
violation, demonstrating why it does not have an easy fix.

\begin{center}
\begin{minipage}{0.48\linewidth}
\includegraphics[width=\linewidth]{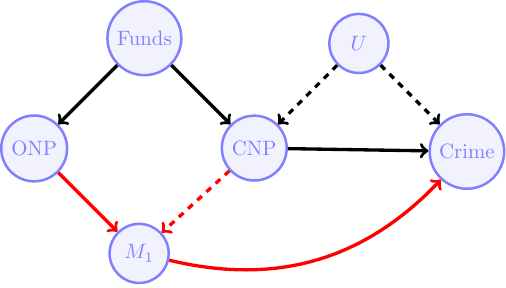}
\captionof*{figure}{\textbf{(A)} An hypothetical mechanism $M_1$ that mediates the (indirect) effect of Other Nonprofits on Crime. For example, building a local community and increasing social capital.}
\end{minipage}%
\hfill
\begin{minipage}{0.48\linewidth}
\includegraphics[width=\linewidth]{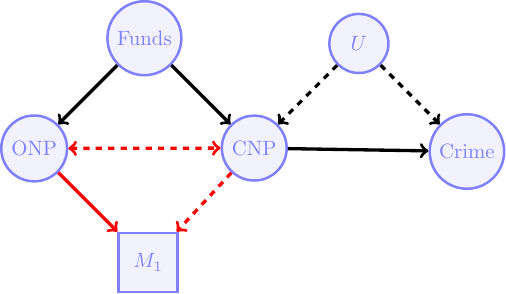}
\captionof*{figure}{\textbf{(B)} If the mechanism is shared between all Community Nonprofit and Other Nonprofits, conditioning on the mechanism do not solve the problem, inducing collider bias.}
\end{minipage}
\captionof{figure}{IV: Possible violations of the exclusion restriction assumption.}
\label{fig:Sharkey_ER}
\end{center}

According to the authors, ``violence is regulated through informal
sources of social control internal to communities.''
\citep[1215]{SharkeyCommunity2017}. These informal sources include,
among others, collective surveillance of public spaces which can
certainly be provided by any type of nonprofit, for example through
``efforts to clean up neighborhoods or improve the physical
infraestructure {[}that{]} can bring more people on to the streets and
increase surveillance of public spaces, in turn reducing criminal
activity.'' Therefore, as shown in \textbf{Panel (A)}, there is a link
between \(ONP\) and this hypothesized mechanism
\(M_1\)\footnote{There are other possible mechanisms too, see for example the following discussion in the literature review section: ``In examining the role of community organizations in crime prevention, Skohan (1988) distinguishes between actions that focus directly on reducing criminal activity in the neighborhood (e.g., requesting more policing or engaging in collective surveillance practiecs) and actions that tackle the underlying social and economic factors that lead to crime (e.g., providing employment opportunities). These crime-reducing efforts emerge from communities' ability to capture problem-solving resources and from the activation of a series of mechanisms of informal internal control'' \citep[p. 1218]{SharkeyCommunity2017}. It is likely that the presence of other nonprofits enhance a community's ability to capture such problem-solving resources.}.

One possible response to this problem could be attempting to measure
this informal community efficacy to then control for it. However, this
has two problems as illustrated in \textbf{Panel (B)}. First, it is
likely that these mechanisms are common to all type of nonprofits. If
this is the case, and there is a link \(CNP \rightarrow M_1\), adjusting
for \(M_1\) would partially block the causal effect of community
nonprofits on crime. A second, more subtle problem, is that being
\(M_1\) a common effect of \(ONP\) and \(CNP\) (a \emph{colllider}),
adjusting for \(M_1\) would distort the association between the two
types of nonprofits, invalidating the instrumental variable
design\footnote{As an additional complication, the authors highlight the weak association founded in previous studies between organizational density and crime decline. The way that they describe this research may suggest that they could be actually trying to estimate a \textit{controlled direct effect} of community nonprofits, net of these shared informal channels. See for example the following passage: ``From a theoretical perspective, many studies use measures of organizational density that include a wide range of establishments and organizations that have no direct relationship to crime and violence. These studies are designed to capture the indirect connection between organizational density and crime through the pathway of informal social control (...) But they are not designed to capture the direct effect of community organizations on crime. Rather than examining all organizations within a neighborhood, we focus on local nonprofit organizations that proliferated in the early 1990s with the specific goal of building stronger communities and confronting the problems of crime and violence.'' However, there is no further mention to this, and the authors do not make attempts to adjust for indirect social control effects.}

\subsubsection{Does Educational Equality Increase Mobility? (Rauscher 2016)}

In ``Does Educational Equality Increase Mobility? Exploiting Nineteenth
U.S. Compulsory Schooling Laws'', \cite{RauscherDoes2016} studies the
causal effect of educational expansion on intergenerational mobility
using linked census data from 1850 to 1930. According to the author,
compulsory school laws were exogenous at the individual level, because
who was affected was determined purely by cohort and state of residency
at the time of the law. Therefore, one can use age as the running
variable in a (fuzzy) regression discontinuity framework to assess how
mobility patterns changed for individuals barely exposed and unexposed
to the policy reform. Other individual characteristics that could
otherwise confound the education-mobility link are assumed to be
randomly distributed in a narrow vicinity of the threshold.

\textbf{Regression discontinuity on time}

Figure \ref{fig:Rauscher} (\textbf{Panel (A)}) illustrates the basic
setting. The author is interested in the relation between social origin
(\(O\), measured as parental ISEI score), individual's education
(\(E\)), and occupational attainment (\(D\)). The effect of education on
mobility is not generally identified, because it is confounded not only
by parental occupational attainment (which can be controlled for), but
also for common factors between parental occupation and children's
education (for example, student's inherited cognitive skills), and
between education and occupational destination (for example, student's
motivation). These unobserved factors are themselved supposed to be
correlated in unknown ways. We can also measure the cohort (\(C\)) an
individual belongs to, which is assumed to correlate with occupational
destination (for example, older people would have more labor market
experience and therefore higher occupational attainment), and secular
time trends (\(T\)) affecting social destination for everyone (through
secular process of economic transformation, including industrialization
and urbanization). Cohort and time trend are allow to be associated.

The basic setting for the RDD design in this case, is to assume that in
a narrow enough time window, 1) cohort is not associated with
destination anymore, and 2) cohort becomes independent of other factors
affecting destination. In other words, Rauscher is making the
assumptions that individuals almost affected and just affected by the
compulsory school law should have the same occupational attainment if
not for education, because they have the same labor market experience
and have been exposed to the same economic conditions. Under this
condition, cohort becomes an instrument for education, allowing the
author to identify the effect of education on destiny. This ``limiting
graph'' \citep{SteinerGraphical2017} is presented in \textbf{Panel (B)}.

\begin{center}
\begin{minipage}{0.48\linewidth}
\includegraphics[width=\linewidth]{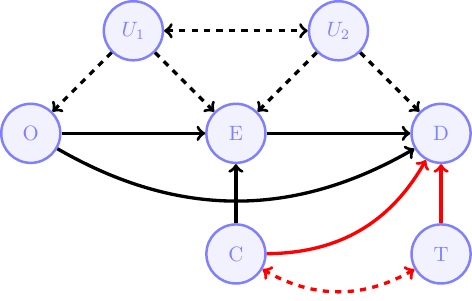}
\captionof*{figure}{\textbf{(A)} Data generating graph: Cohort (C) is associated with educational attainment (E) through educational expansion, with attained socioeconomic status (D) through labor market experience, and other social and economic time trends (T) that affecting mobility.}
\end{minipage}%
\hfill
\begin{minipage}{0.48\linewidth}
\includegraphics[width=\linewidth]{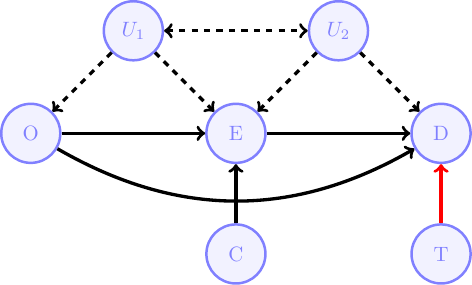}
\captionof*{figure}{\textbf{(B)} Limiting graph: In a narrow window around the cohort cutoff for exposure (C), it is expected that cohort becomes unrelated with social destination (D) and mobility time trends (T), except for its effect on education (E). Cohort becomes a valid IV.}
\end{minipage}
\captionof{figure}{Regression Discontinuity setting from Rauscher (2016)}
\label{fig:Rauscher}
\end{center}

It is important to note here that Rauscher considers as the narrow
window to make the cohort comparisons 10 years at each side of the
cuttof. This make the assumptions just stated very hard to believe. It
is equivalent to compare the occupational attainment of students in the
1980s and the 2000s, which sounds not very realistic. Additionally, one
should consider that, in general, regression discontinuity designs work
better when the running variable is continuos and therefore one can have
people really close to each other to each side. This is not the case
with cohorts, where observations are very sparse and measured in
discrete terms. Some of this issues withe so-called regression
discontinuity on time are discussed in \cite{HausmanRegression2018}. For
now, I will assume that this basic setting holds, to discuss other
potential threats to the validity of this design.

\textbf{Threats to validity}

I will discuss two possible threats to validity in this design. Rauscher
was aware that even in the RDD setting it could be the case that the
assumptions do not perfectly hold. Therefore, the author controls her
estimates including variables in \(T\) to intercept any remaining
association between cohort and destination not channeled through
education. This includes factors like labor market composition (farming,
manufacturing), unemployment, among others. However, and particularly in
the face of the enormous educational transformation induced by the
compulsory laws, it is possible to conjecture that education itself (not
individually, but the educational composition of the labor market, a
shock of high skiller workers) can directly affect the economic
situation. This is expressed in the arrow \(E \rightarrow T\). If this
is the case, adjusting for \(T\) would intercept part of the effect of
education on mobility, biasing the results. This situation is
represented in Figure \ref{fig:Rauscher_descend}, \textbf{Panel (A)}.

Furthermore, if one assumes, as is natural to do, that cohort directly
affect other economic
transformations\footnote{To see why this is the case, note that cohort is perfectly collinear with age and time. Therefore, $C$ also captures economic transformations associated with time, such as migration patterns, inventions, infraestructure projects, etc., all very common at the end of the 19th century.},
the variable \(T\) becomes a collider in the path
\(E \rightarrow T \leftarrow C\). Adjusting for \(T\) would not only
block part of the effect of interest
(\(E \rightarrow T \rightarrow D\)), but also distort the association
between \(T\)'s parents, \(E\) and \(C\), and other ancestors, like the
pair \(O\) and \(C\). This has two implications. First, it will bias the
first stage, the effect of the reform (captured by \(C\)) on education.
Second, it will distort the association between the reform and the
origin-destination link (the direct arrow \(O \rightarrow D\)). This is
a coefficient of interest to Rauscher, because she is trying to identify
how the educational reform affects intergenerational mobility, i.e.,
parent-child correlation in SEI scores. This effects becomes
unidentified due to collider
bias\footnote{This is structurally the same problem discussed by \cite{ZhouEqualization2019}.}.
Figure \ref{fig:Rauscher_descend}, \textbf{Panel (B)}, corresponds to
this issue.

\begin{center}
\begin{minipage}{0.48\linewidth}
\includegraphics[width=\linewidth]{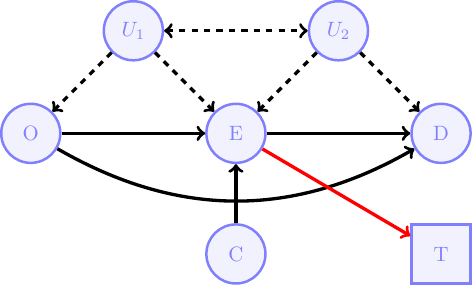}
\captionof*{figure}{\textbf{(A)} If the moblity time trend (T) is a function of educational expansion (E), adjusting for it could remove some of the effect of education on social destination.}
\end{minipage}%
\hfill
\begin{minipage}{0.48\linewidth}
\includegraphics[width=\linewidth]{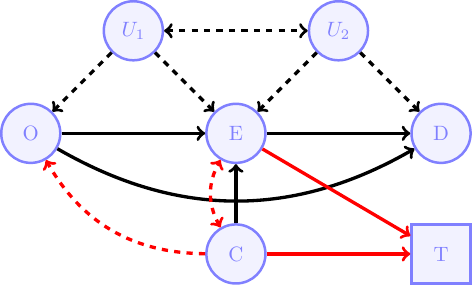}
\captionof*{figure}{\textbf{(B)} If mobility time trend (T) is also a function of cohort characteristics (C), adjusting for it would distort the association of social origin and destination.}
\end{minipage}
\captionof{figure}{Overcontrolling will block some of the effect of education on mobility and distort the associations between social origin and destination by cohort.}
\label{fig:Rauscher_descend}
\end{center}

Finally, an additional point of concern is represented in Figure
\ref{fig:Rauscher_measure}. This is a different problem from what we
have been discussing and remains even assuming the original RDD
assumptions hold within the 10 years window. In this DAG, I incorporate
a new variable \(D*\), which corresponds to children's occupational
attainment measured in a more distant point of time. If there are
\emph{intragenerational} occupational changes, then it is critical that
the measures of destination taken for the exposed and unexposed group
(the children that experienced the compulsory laws and those who did
not) correspond to the same time. However, given the data structure,
this is not possible, with time windows from measure one (social origin)
to measure two (social destination) ranging from 10 to 50 years. The
path \(C \rightarrow D*\) biases the results if measures are taken for
some observations in \(D\) and others in \(D*\). This seems to be the
case, as suggested by the mismatch between results in Table 5 (full
sample) and Table 6 (limited to 30-64 years old during time 2) of the
original paper. This highlights that, even in the presence of a valid
identification strategy, measurement and data issues can substantially
bias the results. Graphical models are useful tools to uncover these
subtle issues.

\begin{center}
\includegraphics{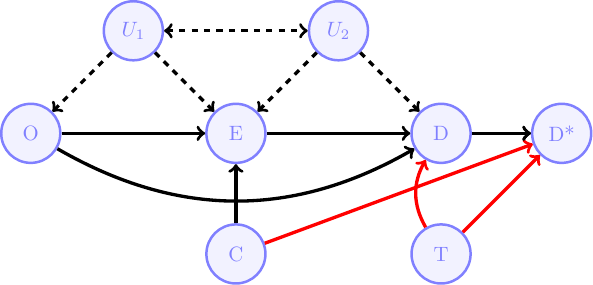}
\captionof{figure}{If social destination (D) is measured at different points in time (D*) depending on the cohort, then cohort will not be a valid instrument anymore.}
\label{fig:Rauscher_measure}
\end{center}

\section{Conclusion: Taking advantage of graphical expressivity}

\begin{quote}
\emph{About 20 years ago, when asked in a meeting what can be done in
observational studies to clarify the step from association to causation,
Sir Ronald Fisher replied: `Make your theories elaborate'. The reply
puzzled me at first, since by Occam's razor, the advice usually given is
to make theories as simple as is consistent with known data. What Sir
Ronald meant, as subsequent discussion showed, was that when
constructing a causal hypothesis one should envisage as many different
consequences of its truth as possible, and plan observational studies to
discover whether each of these is found to hold} \hfill --- B.G. Cochran
(cited by Rosenbaum, 1995)
\end{quote}

In empirical research, models do not need to be externally imposed: they
naturally grow from the substrate of each particular study once we look
carefully enough. Graphical models are just a convenient way to make
explicit and formalize those naturally occurring models, and expose them
to closer examination. If, as I argued, empirical research lies
somewhere in between the ideal and simplest templates, and intractable
complex (and hardly credible) models, then there is ample room for the
expressive capabilities of graphical models to be exploited.

Currently, research communities doing causal inference across the social
sciences seem to be split in two groups: those who exclusively resort to
potential outcomes to formalize their analyses, usually relying on
pre-defined identification strategies or templates, and those who
utilize graphical models to encode qualitative assumptions that (aim to)
justify the analytic approach adopted, usually developing totally
general models of the full data generating process. Despite many efforts
and some progress in this direction, we still have a long way to go
before researchers adopt a truly bilingual and cross-disciplinary
approach to causal inference.

Users of both potential outcomes and graphical models have good reasons
to be cautious about either framework. Some researchers prefer sticking
to quasi-experimental settings, to avoid going astray with implausible
conditioning strategies; others prefer starting by making explicit their
assumptions about the data generating process, perhaps avoiding
situations where the causal effect of interest can only be identified
for a small subset of the population. Based on a series of examples, I
have argued that the usefulness of graphical models is not necessarily
opposed to quasi-experimental or design-based approaches. Relying on the
concept of \emph{evidence generating process}, we can see that neither
observational studies require modeling the ``full graph'', nor
quasi-experiments totally bypass the need for modeling.

I argued that using graphical models as described in the proposal of
building DAGs bottom-up (Table \ref{tab:step_by_step}), can help us to
improve the transparency of our inferences, the testability of our
theories, and the generativity of our scientific discussions. I will
conclude by briefly referring to each point.

In terms of \emph{transparency}, DAGs can become the conceptual
equivalent to sharing the code of quantitative analyses: it is usually
impossible to provide sufficient detail of all decisions that go into
producing a given empirical result, and therefore there is no substitute
to having access to the code that \emph{actually} produced the result.
Similarly, as recently shown by \citet{LundbergWhat2021}, it is many
times unclear what are authors \emph{trying} to estimate, let alone
under what assumptions they claim to do so. Providing a DAG in these
circumstances, rather than being ``superfluous'', is critical to
understand how the authors see the mapping between their particular
setting and the template they are invoking, how identification works out
in their particular circumstances, and what changes can make it break
down.

This bring us to the second point, that of \emph{testability}. Graphical
models provide us with a method, based on \(d-\)separation, to
\emph{systematically} derive implications from a model, thus giving us
the tools to test and falsify scientific claims. It is a common joke to
say that ``robustness checks'' are useless since there is no published
paper that has ever failed to pass one. This is suggestive of the risks
of cherry-picking \emph{which} implications are being tested, and
reinforces the need to focus instead on the implications \emph{of what}
are we testing: the underlying model, under which researchers claim to
identify a given causal
effect\footnote{One important take-away from this is the need to treat quasi-experiments with the same level of scrutiny than observational studies, especially with respect to covariate inclusion and sensitivity to potential violations of identifying assumptions. This is especially important in mixed or conditional strategies, where the target template only works under some form of unconfoundedness. Many times, however, that first step remains totally unattended, producing the paradoxical result that the supposedly weakest link of a causal claim is subjected to a lesser evidentiary standard than any analogous observational study would be.}.
The same goes when assessing the credibility of empirical evidence;
rather than assuming a generally incredulous stance, critics should
engage in producing plausible alternative explanations that can be
tested. As economist Claudia Sahm has said, ``be constructive, bring a
model!''.

Finally, DAGs can be instrumental to improve the \emph{generativity} of
our scientific discussions, going beyond the vague skepticism that
sometimes characterizes our disagreements about empirical results. This
is especially important when, like it is the case in causal inference
with observational data, relying on untestable assumptions is
unavoidable. Simply mentioning the \emph{possibility} of unobserved
confounding, for example, without providing a plausible story on how it
can be the case, does not help to move forward our understanding. Using
graphical models is just a formal and transparent way of sharing our
(contested) understanding about the world, which inevitably informs our
analyses. In this context, scientific criticism and theory refinement is
what occurs in the struggle to erase and redraw nodes and arrows, to
incorporate, omit, or break variables into more granular
mechanisms\footnote{Even more, ``complicating the picture'', in the sense of adding new variables and arrows to a DAG, can be a fruitful exercise for theory building, presenting new opportunities for identification, as highlighted by \citet{KnightCausal2013}. A similar argument can be found in \citet{RohrerThinking2018}, where she proposes climate change as an example where the existence of a very plausible mechanism is what increases our confidence in the inference being made. Breaking down nodes and paths into more fine-grained processes helps to clarify the role of hypothetical mechanisms, opening the (front-)door to new identification opportunities and deriving additional testable implications from the model.}.
Graphical models provide a principled tool for causal reasoning that has
ecumenical potential, across disciplines and ways of theorizing, to
guide these conversations.

\bibliographystyle{api-good}
\bibliography{References}  %%% Uncomment this line and comment out the ``thebibliography'' section below to use the external .bib file (using bibtex) .




% 	\bibitem{keshet2016prediction}
% 	Keshet, Renato, Alina Maor, and George Kour.
% 	\newblock Prediction-Based, Prioritized Market-Share Insight Extraction.
% 	\newblock In {\em Advanced Data Mining and Applications (ADMA), 2016 12th International 
%                       Conference of}, pages 81--94,2016.

% \end{thebibliography}

\end{document}